\documentclass[aps,pre,twocolumn,groupedaddress]{revtex4-1}

\usepackage{amsmath} 
\usepackage{amssymb}  
\usepackage{color}

\usepackage[all]{xy}

\usepackage{accents}

\DeclareSymbolFont{bbold}{U}{bbold}{m}{n}
\DeclareSymbolFontAlphabet{\mathbbold}{bbold}
\newcommand{\onev}{\mathbbold{1}}
\usepackage{calrsfs}
\DeclareMathAlphabet{\pazocal}{OMS}{zplm}{m}{n}
\renewcommand{\mathcal}[1]{\pazocal{#1}}

\newcommand{\norm}[1]{\lVert #1 \rVert}

\usepackage{mathtools}

\newcommand{\myheight}{.213\linewidth}
\newcommand{\myonecolwidth}{.85\linewidth}
\newcommand{\myindent}{\hspace{.35cm}}
\newcommand{\myput}[1]{\put(-158,107){{#1}}}
\renewcommand{\myput}[1]{\put(-91.75,-12){{#1}}}
\newcommand{\myspacebetweenfigures}{\hspace{.01\linewidth}}

\begin{document}

\title{Delayed bet-hedging resilience strategies under environmental fluctuations}



\author{Masaki Ogura}
\email{oguram@is.naist.jp}
\affiliation{Graduate School of Information Science,  Nara Institute of Science and Technology, Ikoma, Nara 630-0192, Japan}

\author{Masashi Wakaiki}
\email{wakaiki@ruby.kobe-u.ac.jp}
\affiliation{Graduate School of System Informatics, Kobe University, 
Nada, Kobe, Hyogo 657-8501, Japan}

\author{Harvey Rubin}
\email{rubinh@upenn.edu}
\affiliation{Department of Medicine, University of Pennsylvania, Philadelphia, PA 19104, USA}

\author{Victor M.~Preciado}
\email{preciado@seas.upenn.edu}
\affiliation{Department of Electrical and Systems Engineering, University of Pennsylvania, Philadelphia, PA 19104, USA}

\date{\today}
 
\begin{abstract}
Many biological populations, such as bacterial colonies, have developed through
evolution a protection mechanism, called \emph{bet-hedging}, to increase their
probability of survival under stressful environmental fluctutation. In this
context, the concept of \emph{preadaptation} refers to a common type of
bet-hedging protection strategy in which a relatively small number of
individuals in a population stochastically switch their phenotypes to a
`dormant' metabolic state in which they increase their probability of survival
against potential environmental shocks. Hence, if an environmental shock took
place at some point in time, preadapted organisms would be better adapted to
survive and \emph{proliferate} once the shock is over. In many biological
populations, the mechanisms of preadaptation and proliferation present delays
whose influence in the fitness of the population are not well-understood. In
this paper, we propose a rigorous mathematical framework to analyze the role of
delays in both preadaptation and proliferation mechanisms in the survival of
biological populations, with an emphasis on bacterial colonies. Our theoretical
framework allows us to analytically quantify the average growth rate of a
bet-hedging bacterial colony with stochastically delayed reactions with
arbitrary precision. We verify the accuracy of the proposed method by numerical
simulations and conclude that the growth rate of a bet-hedging population shows
a non-trivial dependency on their preadaptation and proliferation delays.
Contrary to the current belief, our results show that faster reactions do not,
in general, increase the overall fitness of a biological population.
\end{abstract}


\maketitle

\section{Introduction}

Most biological populations are exposed to environmental fluctuations, from
daily regular cycles of light and temperature to irregular fluctuations of
nutrients and pH levels~\cite{Kussell2005,Acar2008}. In this context, many
biological populations employ a protection mechanism called
\emph{bet-hedging}~\cite{Seger1987} to increase their robustness against
potential environmental fluctuations. An important type of bet-hedging
mechanisms, which we call \emph{preadaptation}, is the phenomenon in which the
population `bets' against the presence of prolonged favorable environmental
conditions by having a small number of individuals behaving as if they sensed a
threatening or stressful environment. In the context of bacterial colonies, a
small number of individuals in a population preadapt to environmental shocks by
stochastically 
switching their phenotypes to a `dormant' metabolic state in
which they exhibit slower growth but higher resilience against environmental
shocks, such as antibiotics or pH changes. Hence, if an environmental shock
takes place at some point in the future, we can expect that preadapted
individuals would be better adapted to survive and \emph{proliferate},
rebuilding the bacterial colony once the shock is over.

For example, in the population of \emph{Escherichia coli} on a mixture of
glucose and lactose, it has been observed that the population typically contains
a small portion of individuals activating the \emph{lac} operon for consuming
lactose, despite the fact that glucose is much easier to digest and leads to
higher growth rates~\cite{Boulineau2013}. As a result, the population as a whole
increases its chances to survive through a sudden lack of glucose, while
sacrificing short-term performance. Similar bet-hedging strategies can be found
in many other biological systems, such as lysis-lysogeny switch of
bacteriophage~$\lambda$~\cite{Oppenheim2005}, delayed germination in
plants~\cite{Gremer2014}, and phenotypic variations in
bacteria~\cite{Woude2004}.

Due to the importance of bet-hedging mechanisms in biological populations, we
find in the literature various analytical
tools~\cite{Belete2015,Kussell2005,Gaal2010,Muller2013a,Belete2015,Skanata2016,Sughiyama2017} to quantify the growth rates of bet-hedging populations under fluctuating environments. In many real populations, we find time delays associated to bet-hedging mechanisms. For example, while studying the growth of \emph{Escherichia coli} on a mixture of glucose and lactose, the authors in~\cite{Boulineau2013} found stochastic delays in the activation of the \emph{lac} operon by individual cells in response to their exposition to lactose-only mediums. In the context of plant populations, delayed germination~\cite{Gremer2014} and delayed disease activation of viruses~\cite{Stumpf2002} have also been reported as bet-hedging strategies. Furthermore, the presence of time-delays in some basic patterns of cell proliferation is known to improve the overall population fitness~\cite{Baker1998}. Despite the common presence of time delays in bet-hedging mechanisms, current analytical tools~\cite{Belete2015,Kussell2005,Gaal2010,Muller2013a,Belete2015,Skanata2016,Sughiyama2017} neglect these delays for the sake of simplicity, failing to analyze the effect of delays on the growth rates of biological populations.

In this paper, we present a rigorous and tractable framework to
quantify the growth rates of cell populations employing bet-hedging
strategies subject to time-delays. Building on current delay-free models~\cite{Thattai2004,Kussell2005,Belete2015}, we introduce dynamical models of bet-hedging populations subject to
delays by using stochastic differential equations.
Among various types of delays, we specifically focus on those present in both proliferation and
preadaptation mechanisms.
The accuracy of our theoretical results are confirmed via numerical simulations.
Our analysis shows that the growth rates of bet-hedging populations subject to delays depend in a highly nontrivial way on the delays. In contrast to current belief, we show that the shorter delays would not, in general, increase the overall fitness of populations.

\section{Delayed Proliferations}
\label{sec:delayedProliferation}

\begin{figure}[tb]
\centering
\includegraphics[width=\linewidth]{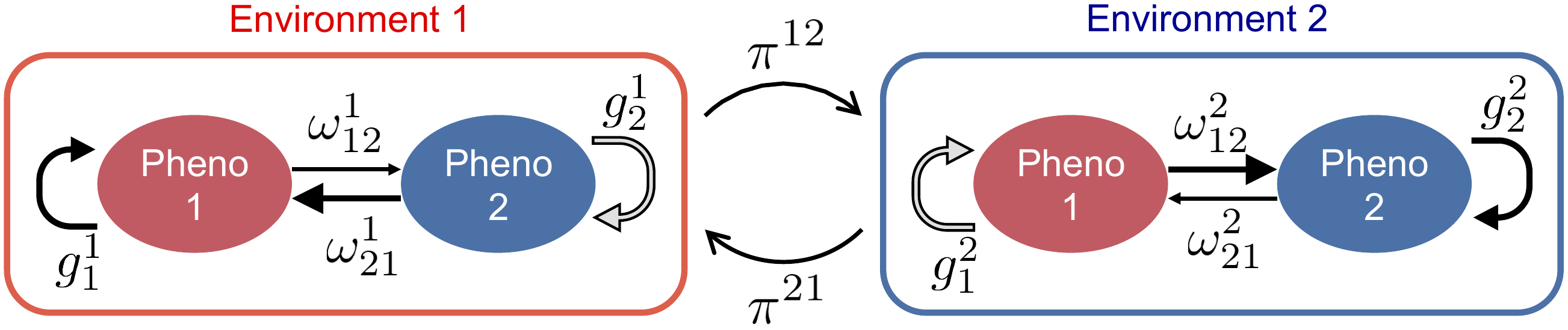}
\caption{Bet-hedging population ($n=2$).}
\label{fig:offset}
\end{figure}

We consider a bet-hedging biological population growing in an environment that
fluctuates in time among $n$ possible environmental types. The fluctuation is
modeled~\cite{Kussell2005} by a continuous-time Markov process~$\epsilon(t)$
taking values in $\{1,\dotsc,n\}$, where $\epsilon(t)$ designates which
environment occurs at time $t$. The infinitesimal generator of the Markov
process $\epsilon(t)$ is the $n\times n$ matrix $\Pi = [\pi^{ij}]_{i,j}$ whose
entries represent the transition rates between environments (throughout the
paper, we use superscripts to denote environmental variables). The probability
of an environmental transition taking place during a time-window of length~$h$
is, hence, given by
\begin{equation*}
P(\epsilon(t+h) = j \mid \epsilon(t) = i) = 
\begin{cases}
\pi^{ij} h + o(h),		&\text{if }j\neq i,\\
1+\pi^{ii} h + o(h),\!\!	&\text{if }j=i, 
\end{cases}
\end{equation*}
where $o(h)$ denotes a term of small order in $h$ (i.e., $o(h)/h\to 0$ as $h\to 0$). We assume that, in response to environmental fluctuations, each individual in the population can
exhibit one of $n$ different phenotypes, denoted by $1$, $\dotsc$, $n$ (throughout the paper, we shall use subscripts to denote phenotypic variables). Under environment~$i$, we denote the instantaneous growth rate of those individuals with phenotype~$k$ by $g_k^i$.  By convention, we assume that phenotype $i$ presents the largest growth rate in environment $i$ (i.e., $g_i^i\geq g_k^i$). 

In a bet-hedging population, individuals may stochastically switch their phenotype at any time. We denote by~$\omega_{k\ell}^i$ the instantaneous rate at which an individual having phenotype~$k$ adaptively switches its phenotype to phenotype~$\ell$ under environment~$i$. The number of individuals having
phenotype~$k$ at time~$t$ in the population is denoted by $x_k(t)$. The dynamics of the population can be
modeled by the following set of differential equation~\cite{Kussell2005,Belete2015}
\begin{equation}\label{eq:baseModel}
\frac{dx_k}{dt}=g_k^{\epsilon(t)}x_k(t)
+\sum_{\ell= 1}^n \omega_{\ell k}^{\epsilon(t)}x_{\ell}(t),
\end{equation}
where $\omega^i_{kk} = -\sum_{\ell\neq k}\omega_{k\ell}^i$. See
Fig.~\ref{fig:offset} for a schematic picture of this model for $n=2$, i.e., individuals present two types of phenotypes in two possible environments.

Previous studies analyzing bet-hedging strategies neglect the effect of delays in the population growth. In this paper, we analyze the effect of these delays and show how they can induce nontrivial effects in the population growth. We start our analysis by extending the dynamic model in~\eqref{eq:baseModel} to include the effect of delayed proliferation~\cite{Stumpf2002,Baker1998} in  bet-hedging populations, as follows
\begin{equation}\label{eq:dx_k/dt:p_delay}
\frac{dx_k}{dt}=g_k^{\epsilon(t)} x_k(t)
+
\sum_{\ell=1}^n \omega_{\ell k}^{\epsilon(t)}x_{\ell}(t) 
+
p_k^{\epsilon(t)}x_k(t-d^{\epsilon(t)}_k), 
\end{equation}
where $d^i_k$ denotes the proliferation delay of the individuals with
phenotype~$k$ in the environment $i$, and~$p^i_k$ denotes the factor of the
delayed growth. The main objective of this section is to give an analytical
framework to quantify the growth rate~$\rho$ of the size of
the total population~$x_1 + \cdots + x_n$.

In order to quantify the growth rate~$\rho$, it is convenient
to work with vectorial representations. We first introduce a vectorial
representation of the environmental dynamics. For each time $t$, define the
$n$nobreakdash-dimensional vector~$\eta(t) = (\eta^1(t), \dotsc, \eta^n(t))$ by
$\eta^i(t) = 1$ if $\epsilon(t) = i$ and~$\eta^i(t) = 0$ otherwise. It is
known~\cite{Brockett2009} that the dynamics of the variable~$\eta$ can be
described by the following Poisson-type stochastic differential equation
\begin{equation}\label{eq:deta}
d\eta
=
\sum_{i=1}^n \sum_{j\neq i} (U_{ji}-U_{ii})\eta\,dN_{\pi^{ij}}, 
\end{equation}
where $N_{\pi^{ij}}$ denotes a Poisson counter of rate $\pi^{ij}$
and $U_{ij}$ denotes a $0$-$1$ matrix whose entries are all zero
except its $(i, j)$-th entry. In order to express the population
model~\eqref{eq:dx_k/dt:p_delay} in a vectorial form, let us define
the matrix~$A^i =\bigoplus(g_1^i, \dotsc, g_n^i) +
([\omega_{k\ell}^i]_{k,\ell})^\top$ (where $\bigoplus$ denotes the
direct sum of matrices) and the vector variables
\begin{equation}\label{eq:defx}
x = \begin{bmatrix}
x_1\\\vdots\\x_n
\end{bmatrix}, 
\ 
(B^ix)(t) = \begin{bmatrix}
p^i_1 x_1(t-d^i_1)\\\vdots\\ p^i_n x_n(t-d^i_n)
\end{bmatrix}. 
\end{equation}
We can then rewrite \eqref{eq:dx_k/dt:p_delay} as ${dx}/{dt} =
A^{\epsilon(t)} x(t) +(B^{\epsilon(t)}x)(t)$. Introducing the notation
$(\mathcal A^i x)(t) = A^i x(t) + (B^ix)(t)$, we can further obtain 
the simple form~$dx/dt = \mathcal A^{\epsilon(t)}x$. From this
representation and the definition of the environmental variables
$\eta^i$, we finally obtain the following multiplicative stochastic differential
equation for the population dynamics:
\begin{equation}\label{eq:new:Sigma1}
dx=
\sum_{i=1}^n
\eta^i(t) \bigl(
(\mathcal{A}^i x)(t)
\bigr)dt, 
\end{equation}
where the evolution of the environmental variable~$\eta$ is described in
\eqref{eq:deta}. For other examples of multiplicative stochastic
differential equations in the context of biology, we refer the readers to, e.g.,
\cite{Ciuchi1993,Hasty2000}.

In what follows, we quantify the growth rate~$\rho$ of the
total population by studying an auxiliary stochastic process defined by the
following Kronecker product of vectors:
\begin{equation*}
z = \eta \otimes x.
\end{equation*}
Notice that the $1$-norm~$\norm{E[z]}$ of the mathematical
expectation $E[z]$ satisfies $\norm{E[z(t)]} = \sum_{i, k = 1}^n
E[\eta^i(t)x_k(t)] = E[\sum_{k=1}^n x_k(t)]$, due to the positivity
of~$x$ and the obvious identity~$\sum_{i=1}^n \eta^i = 1$. From this
fact, we see that the expected growth rates are given by the (deterministic) vector~$E[z]$. It further turns
out that working with this auxiliary variable $z$ is easier than
directly studying the population vector~$x$, as we will see below. In
order to quantify the growth rate of $E[z]$, we first apply Ito's
Lemma to the stochastic differential equations~\eqref{eq:deta} and
\eqref{eq:new:Sigma1} to obtain (after simple but tedious calculations),
\begin{equation}\label{eq:d(del.ox.x).pre}
\begin{aligned}
dz
=&
\sum_{i=1}^n(\eta\otimes I_n)\eta^i(\mathcal{A}^i x)\,dt 
\\
& \myindent\myindent+\sum_{i=1}^n \sum_{j\neq i}\left[\bigl((U_{ji}-U_{ii})\eta\bigr)\otimes x \right]dN_{\pi^{ij}}, 
\end{aligned}
\end{equation}
where $I_n$ is the identity matrix of dimension~$n$. We can further
show (see the Appendix for more details) that the
expectation $\zeta = E[z]$ obeys the differential equation
\begin{equation}\label{eq:barMJLSD}
\frac{d\zeta}{dt}=\mathbf A_0\zeta(t)
+
\sum_{i,k=1}^n \mathbf A^i_k \zeta(t-d^i_k), 
\end{equation}
where $\mathbf A_0 = \Pi^\top \otimes I_n + \bigoplus (A^1,
\dotsc, A^n)$, $\mathbf A^i_k = p^i_k u_i\otimes (U_{ki}
e^{\Pi^\top d^i_k})\otimes u_k^\top$, and $\{u_1, \dotsc, u_n\}$ is
the canonical basis of the $n$\nobreakdash-dimensional Euclidean
space.

\begin{figure}[tb]
\centering \includegraphics[width=\myonecolwidth]{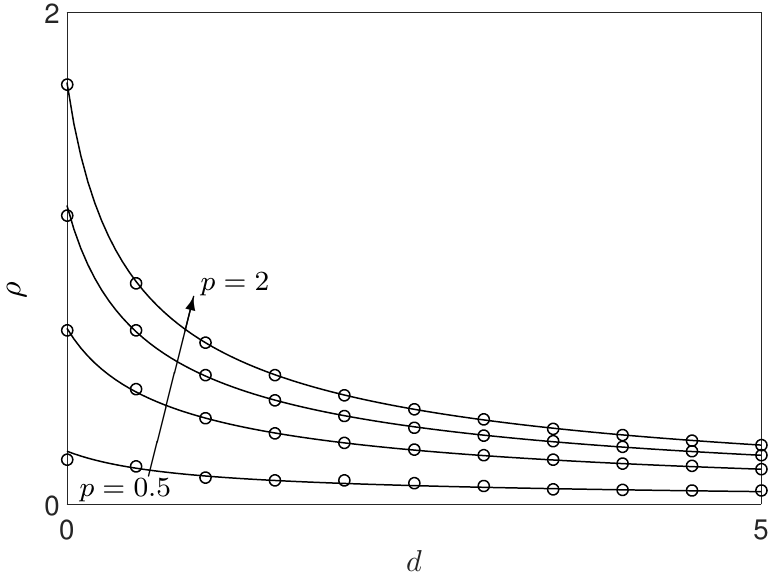}
\caption{Growth rates
$\rho$ of the dynamical population
model~\eqref{eq:dx_k/dt:p_delay} for $p = 0.5$, $1$, $1.5$, and~$2$ (from the
bottom). Solid lines: Growth rates predicted by the proposed method. Circles:
Growth rates estimated from the sample trajectories of the model.}
\label{fig:state:compare}
\end{figure}

\begin{figure*}[tb]
\centering \includegraphics[height=\myheight]{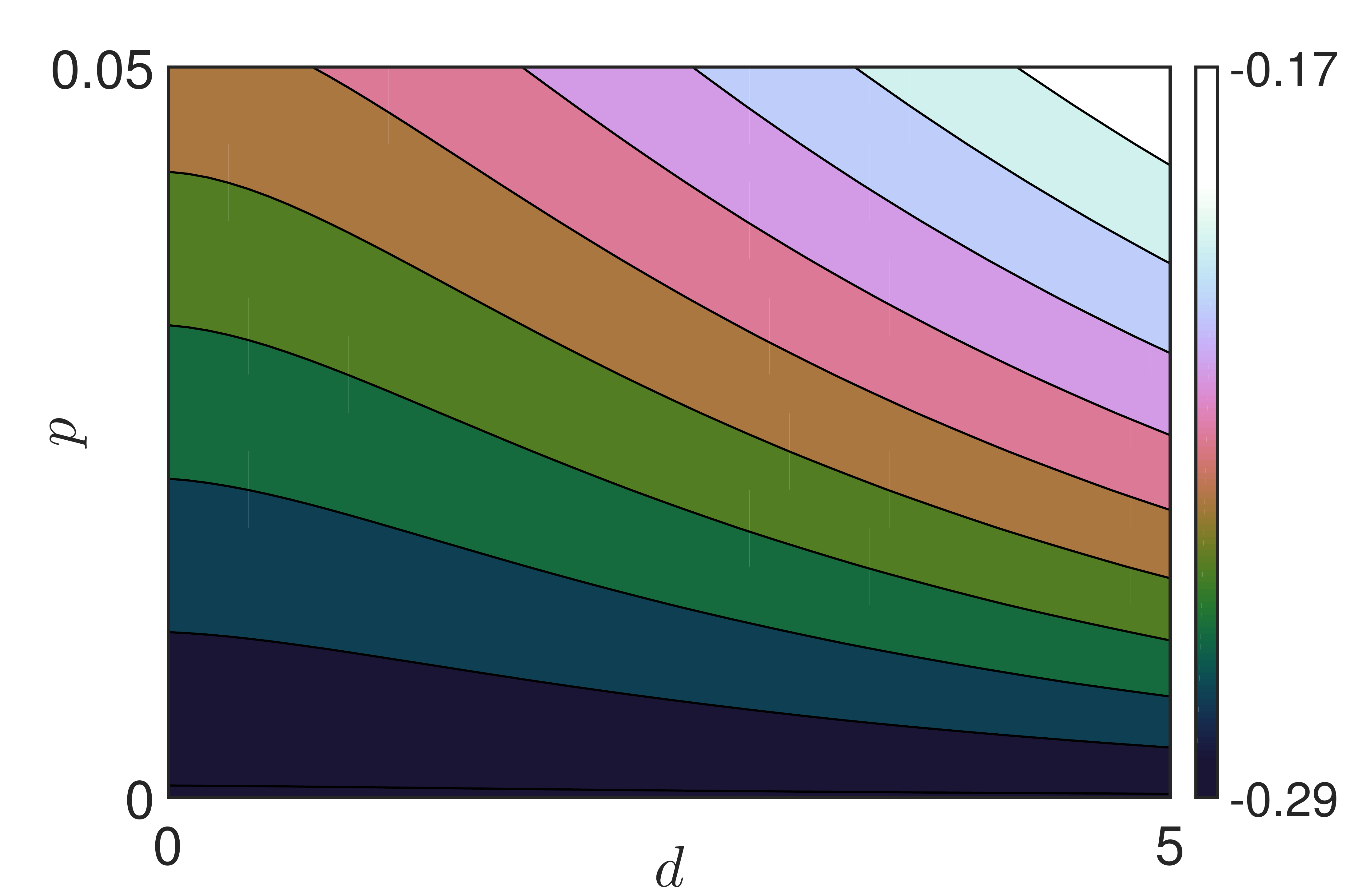} \myput{(a)}
\myspacebetweenfigures \includegraphics[height=\myheight]{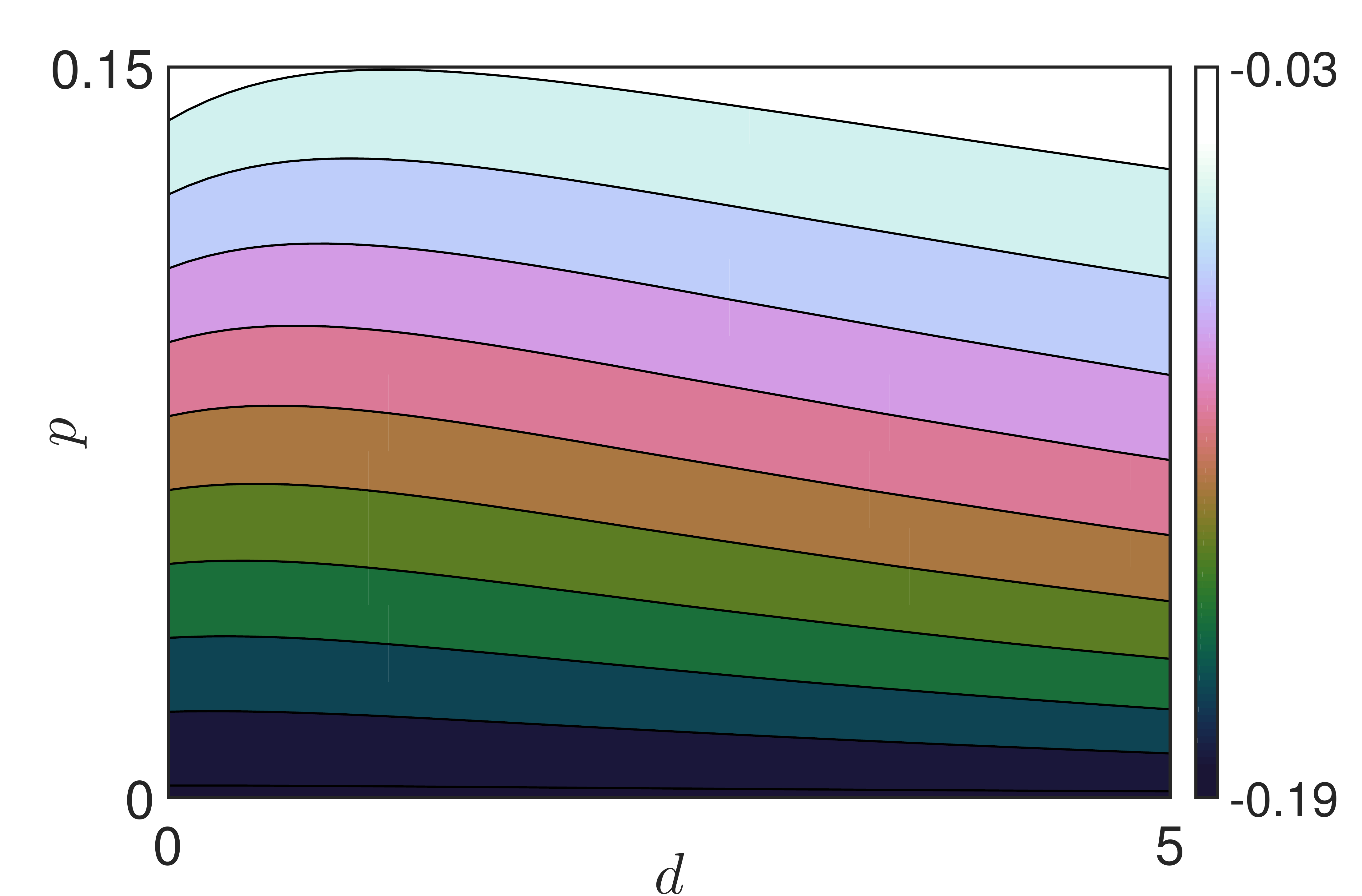} \myput{(b)}
\myspacebetweenfigures \includegraphics[height=\myheight]{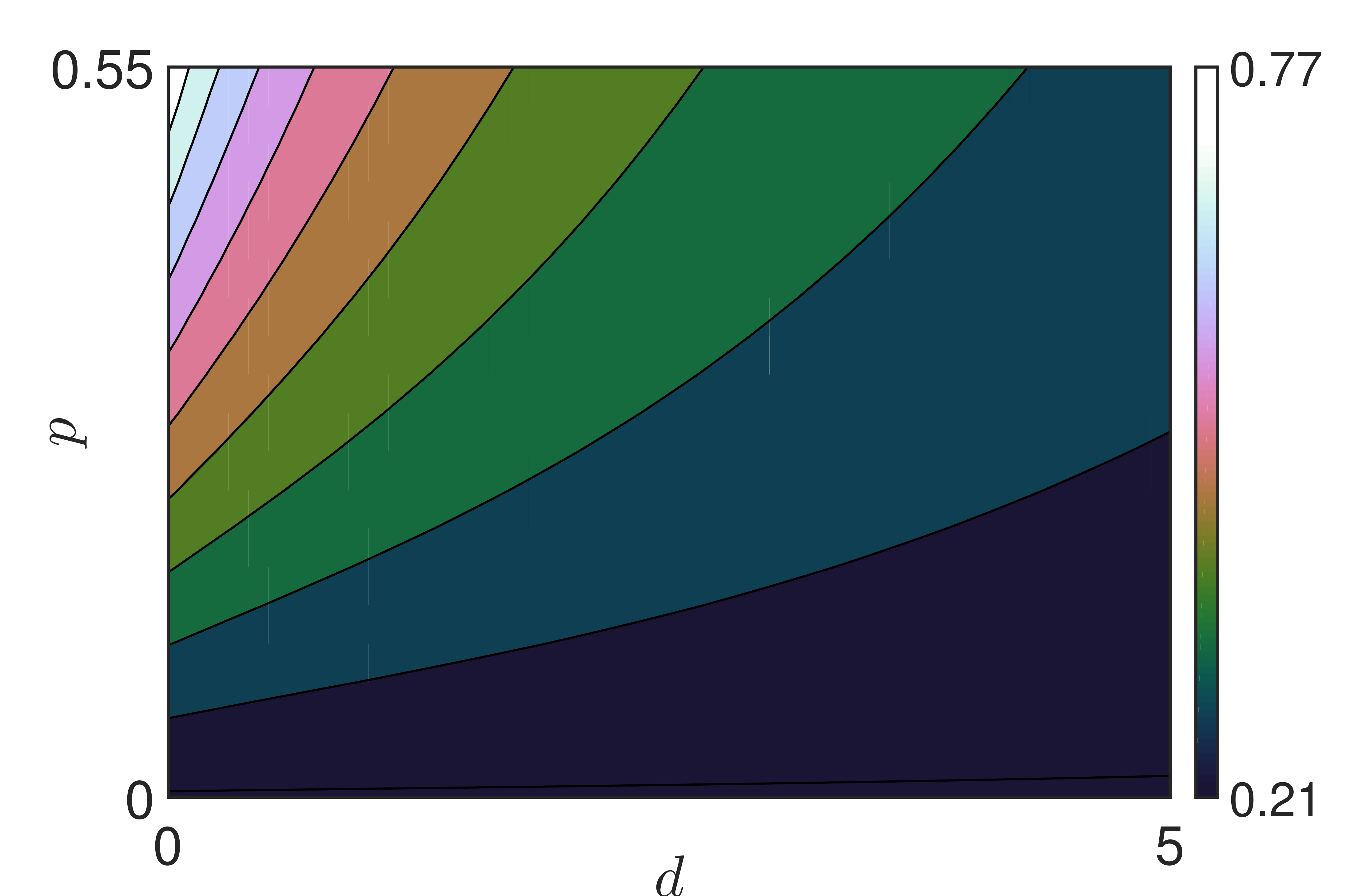} \myput{(c)}
\caption{Growth rates $\rho$ versus $d$ and~$p$ for various global fitness
parameter~$\tau$: (a) $\tau = 0$, (b) $\tau = 0.1$, (c) $\tau = 0.5$.
The figures show a nontrivial dependency of the growth rates on
the parameters $d$, $p$, and $\tau$. Specifically, when $\tau = 0$
(Fig.~\ref{fig:state:p d}(a)), we observe that, the larger the delay, the higher
the growth rates; while for $\tau = 0.5$ (Fig.~\ref{fig:state:p d}(c)), we
observe the opposite tendency.} \label{fig:state:p d}
\end{figure*}

\begin{figure*}[tb]
\centering \includegraphics[height=\myheight]{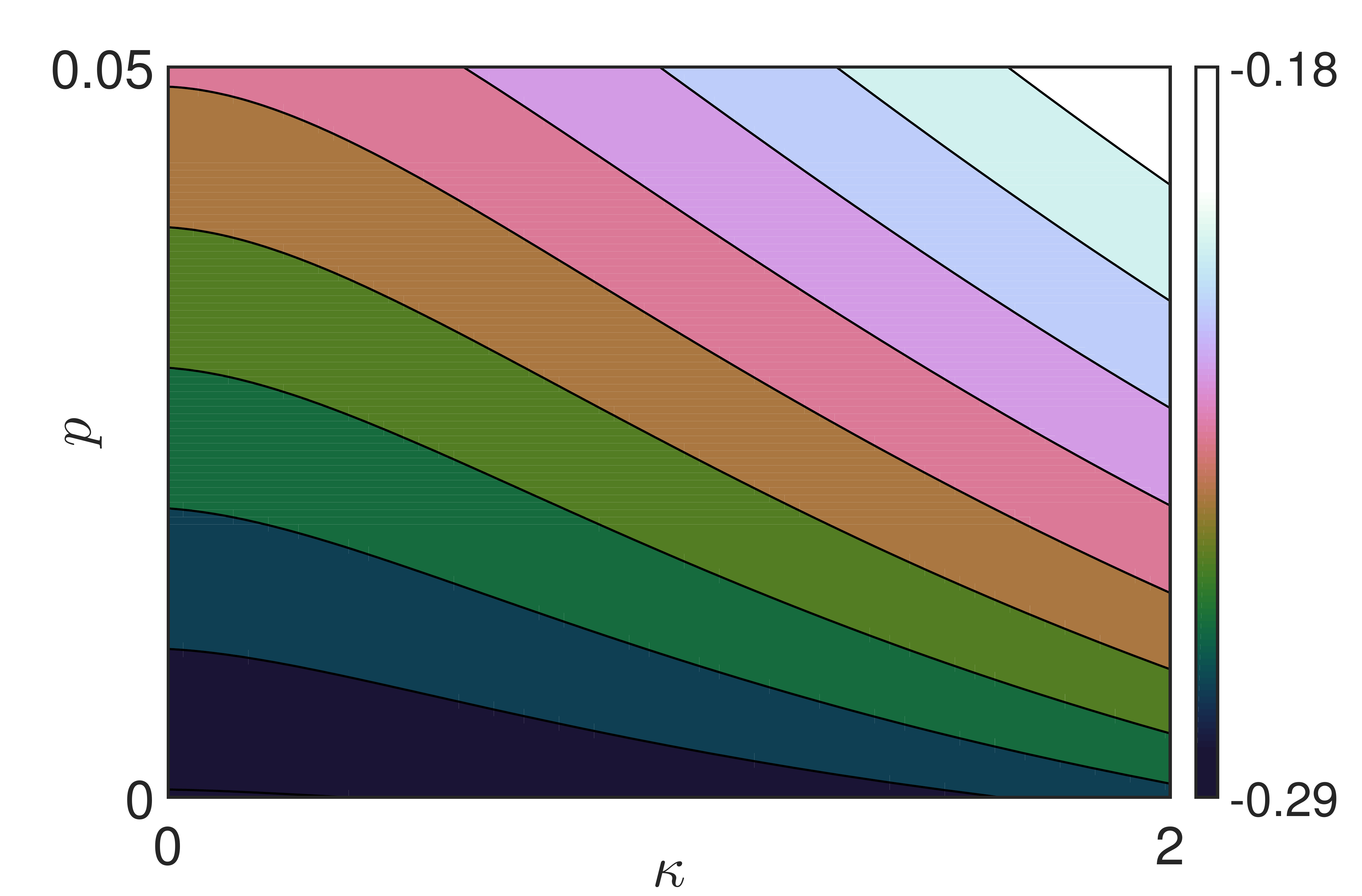} \myput{(a)}
\myspacebetweenfigures \includegraphics[height=\myheight]{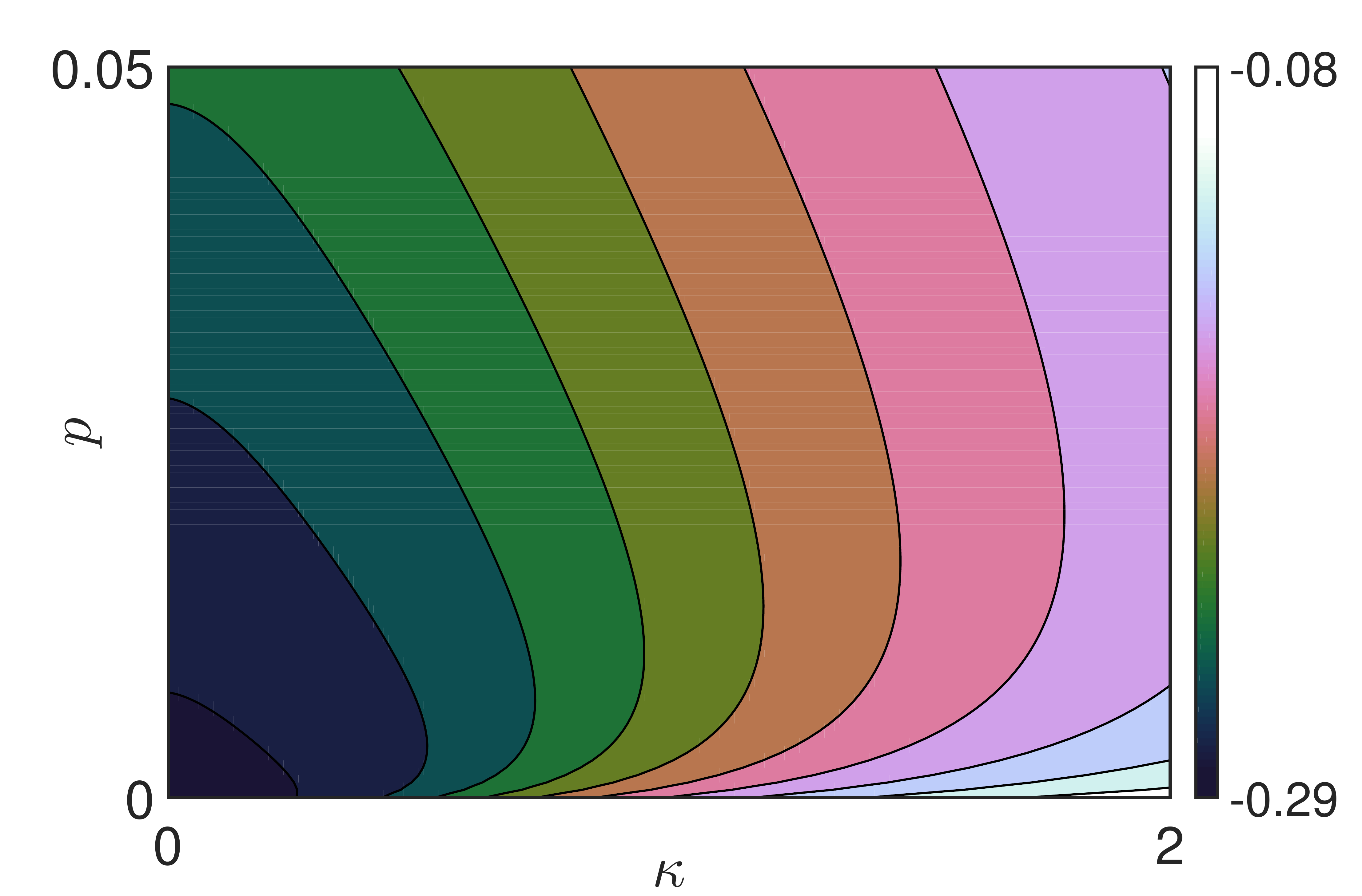} \myput{(b)}
\myspacebetweenfigures \includegraphics[height=\myheight]{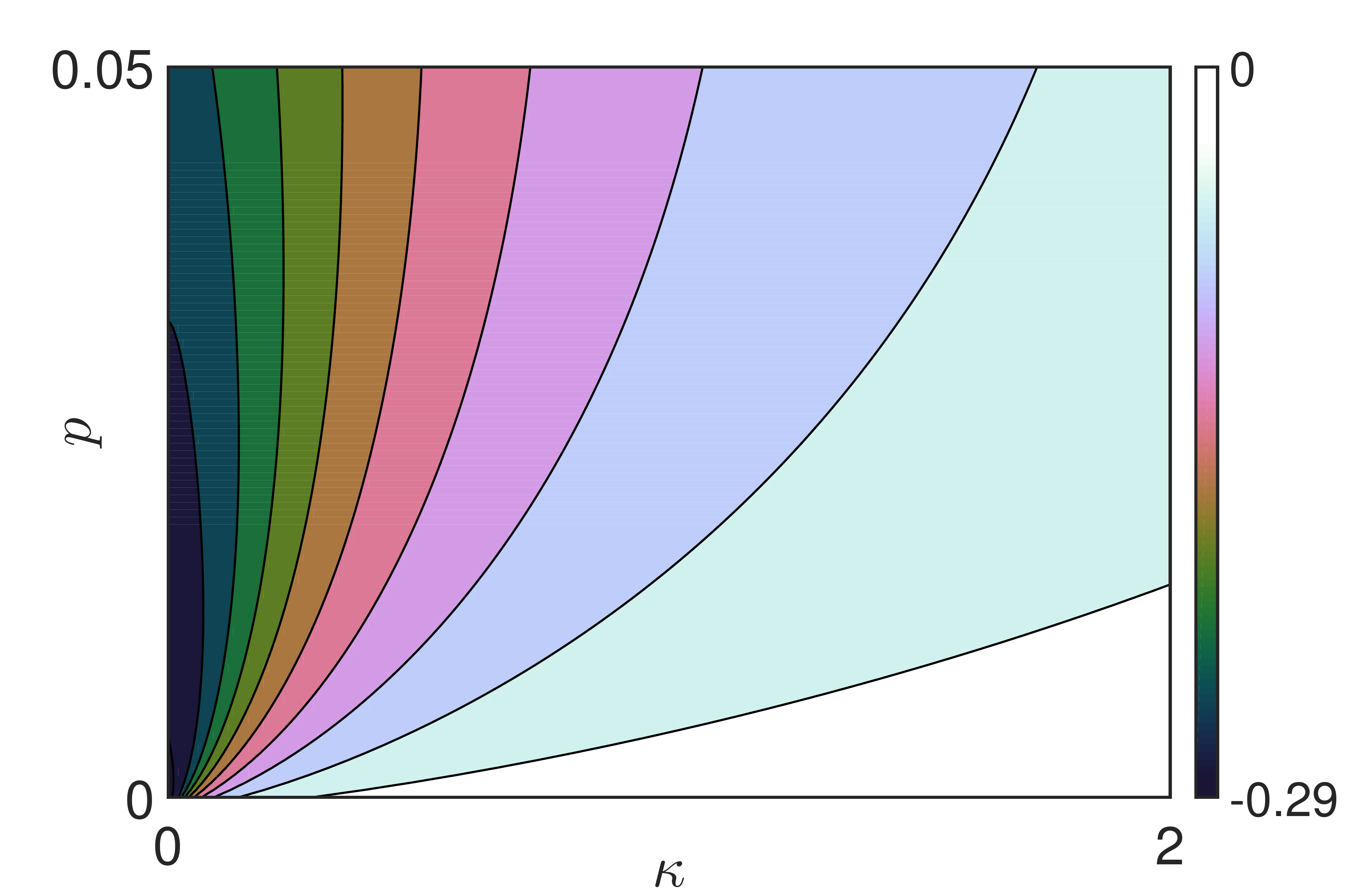} \myput{(c)}
\caption{Growth rates $\rho$ versus $\kappa$ and~$p$ for various exponent~$\phi$ on $p$: (a) $\phi = 0.25$, (b) $\phi =
0.5$, (c) $\phi = 0.95$. The figures show that the exponent~$\phi$  drastically changes the dependency of the growth rates on
the delayed growth factor~$p$. For example, in the regime of small $\phi$
(Fig.~\ref{fig:state:pExponent}(a)), we observe that a good strategy for a
population to increase its growth rate is to bet on a delayed growth factor~$p$.
On the other hand, in the case of relatively large $\phi$
(Fig.~\ref{fig:state:pExponent}(c)), we see that keeping $p$ small results in a higher growth rates.} \label{fig:state:pExponent}
\end{figure*}

Notice that, in \eqref{eq:barMJLSD}, the matrix~$\mathbf A_0$
is a Metzler matrix (i.e., it has nonnegative off-diagonals) and the matrices $\mathbf
A_k^i$ are all nonnegative
entry-wise. Therefore, a standard result from control
theory~\cite[Theorem~III.1]{Ngoc2013} shows that, the real parts of
the eigenvalues of the sum of the matrices in the
differential equation~\eqref{eq:barMJLSD} (that is, the matrix~$\mathbf
A_0 + \sum_{i,k=1}^n \mathbf A_k^i$) are all negative, if and only if, the
growth rate of the vector $\zeta$ (and, hence, the growth rate~$\rho$ of the total
population) is negative. In order to obtain more precise information
on the growth rate, we can work with the auxiliary variable $\tilde
x(t) = e^{-\lambda t}x(t)$ for a real parameter~$\lambda$. Observe
that, by \eqref{eq:dx_k/dt:p_delay}, this variable~$\tilde x$
satisfies the differential equation
\begin{equation*}\label{eq:dx/dt^lambda}
\begin{aligned}
\frac{d\tilde x_k}{dt}=&(g_k^{\epsilon(t)}-\lambda) \tilde x_k(t)
\\
&\myindent+\sum_{\ell=1}^n \omega_{\ell k}^{\epsilon(t)}\tilde x_{\ell}(t) 
+
e^{-\lambda d^{\epsilon(t)}_k}p_k^{\epsilon(t)}\tilde x_k(t-d^{\epsilon(t)}_k). 
\end{aligned}
\end{equation*}
In the same way as we derived \eqref{eq:barMJLSD}, we see that the
variable $\tilde \zeta = E[\eta \otimes \tilde x]$ satisfies the
differential equation ${d\tilde \zeta}/{dt}=(\mathbf A_0 - \lambda
I_n)\tilde \zeta(t) + \sum_{i,k=1}^n \tilde {\mathbf
A}_k^i\tilde{\zeta}(t-d_k^i)$, where $\tilde{\mathbf A}_k^i = p_k^i
e^{-\lambda d_k^i} u_i\otimes (U_{ki} e^{\Pi^\top d^i_k})\otimes
u_k^\top$. Therefore, again by \cite{Ngoc2013}, the real parts of the
eigenvalues of the matrix
\begin{equation*}
\mathbf A_0-\lambda I_n
+
\sum_{i,k=1}^n \tilde {\mathbf A}_k^i, 
\end{equation*}
are all negative, if and only if, the growth rate of $E[\tilde \zeta]$
is negative; or, equivalently, the growth rate~$\rho$ of the total population
size is less than $\lambda$. This result allows us to efficiently
quantify the growth rate of the total population within an arbitrary
accuracy by employing a simple bisection search on the parameter $\lambda$.

\subsection{Numerical simulations}

In this section, we illustrate our results via several simulations. For
clarity in our presentation, we focus on the case where there are two
possible phenotypes and two possible environments ($n=2$). The parameters of the delay-free model~\eqref{eq:baseModel} are given by:
\begin{equation*}
g_1^1=0.05+\tau,\ g_2^1 = \tau,\ 
g_1^2=-10+\tau, \ g_2^2 = -0.1+\tau, 
\end{equation*}
where $\tau$ is a real parameter that we can tune to adjust the global fitness of individuals. Observe how, following our convention, phenotypes 1 and 2 are better fitted
to the environments~1 and~2, respectively. In our simulations, we fix the phenotypic
transition rates as $\omega^1_{12}=\omega^2_{21} = 0.1$ and
$\omega^1_{21} = \omega^2_{12}= 1$. The rates of environmental changes
are chosen as $\pi^{12} = 0.5$ and~$\pi^{21} = 1$. For simplicity in
our analysis, we use homogeneous values for the delays and rates of delayed
proliferation, i.e., we let $d^i_k = d$ and~$p^i_k = p$ for all
environments~$i$ and phenotypes~$k$. With the above parameters, we run a simulation to
validate the accuracy of the proposed method.
Fig.~\ref{fig:state:compare} shows the actual growth rates of the
total population (circles) and the growth rates computed by the
proposed method (solid lines). In order to compute the actual growth
rates, we compute 1,000 sample paths of the dynamics~\eqref{eq:dx_k/dt:p_delay} for each pair of parameters and then fit
an exponential functions to the sample average of the population size. Observe how the predictions from our analysis provide accurate estimates for the actual growth rates of the population.

We now analyze the dependency of the growth rate on the parameters~$p$ (the
delayed factor growth) and~$d$ (the delay length), as shown in
Fig.~\ref{fig:state:p d}.  These figures show a nontrivial dependency of the
growth rates on these parameters, as well as the global fitness
parameter~$\tau$. When $\tau = 0$ (Fig.~\ref{fig:state:p d}(a)), we observe
that, the larger the delay, the higher the growth rates; while for $\tau = 0.5$
(Fig.~\ref{fig:state:p d}(c)), we observe the opposite tendency. In the
intermediate regime (Fig.~\ref{fig:state:p d}(b)), we observe how the growth
rates are not very sensitive to delays. Intuitively speaking, these observations
show that, in harsh environments, it is better for individuals to have delays in
their proliferations; while in favorable environments, immediate reproductions
would be a better strategy to increase the overall fitness of the population.

In our simulations, we also investigate the trade-off between the delay length~$d$ and the delayed growth factor~$p$. This kind of trade-off is observed in, for example, the case of diauxic growth of \emph{Escherichia coli} in a
medium containing glucose and lactose~\cite{Chu2016}. For our
analysis, we consider the situation in which the population can `tune'
the parameters $p$ and $d$ under the simple constraint~$p^\phi d^\psi
= \kappa$, where $\phi$, $\psi$, and~$\kappa$ are positive parameters
that would be specific to biological populations. We shall fix $\psi =
1$ without loss of generality. For $\phi = 0.25$, $0.5$, and~$0.95$,
we show the growth rates of the population versus $\kappa$ and~$p$ in
Fig.~\ref{fig:state:pExponent}. We can observe that the
exponent~$\phi$ on $p$ drastically changes the dependency of the
growth rates on the delayed growth factor~$p$. For example, in the
regime of small $\phi$ (Fig.~\ref{fig:state:pExponent}(a)), we observe
that a good strategy for a population to increase its growth rate is to bet on a delayed growth factor~$p$. On the other hand, in the case of
relatively large $\phi$ (Fig.~\ref{fig:state:pExponent}(c)), we see
that keeping $p$ small (which implies a longer delay in
proliferation) results in a higher growth rates. This observation
suggests that the trade-off between the delay length and the delayed
growth factor is not trivial.

\section{Delayed adaptations} \label{sec:delayedAdaptation}

Besides the delays in proliferations studied in
Section~\ref{sec:delayedProliferation}, bet-hedging populations can
experience delays in their \emph{adaptation} to, as well as \emph{sensing} of,
environmental fluctuations. For example, experiments on
single cells have recently shown~\cite{Boulineau2013} that
it can take several hours for the individual cells in the colony of
\emph{Escherichia coli} to phenotypically adapt to nutritional changes
of their surrounding environment. The aim of this section is to
propose a dynamical model of bet-hedging populations having such
adaptation delay and derive a tractable framework for exactly
quantifying their growth rates.

\begin{figure}[tb]
\centering \includegraphics[width=\linewidth]{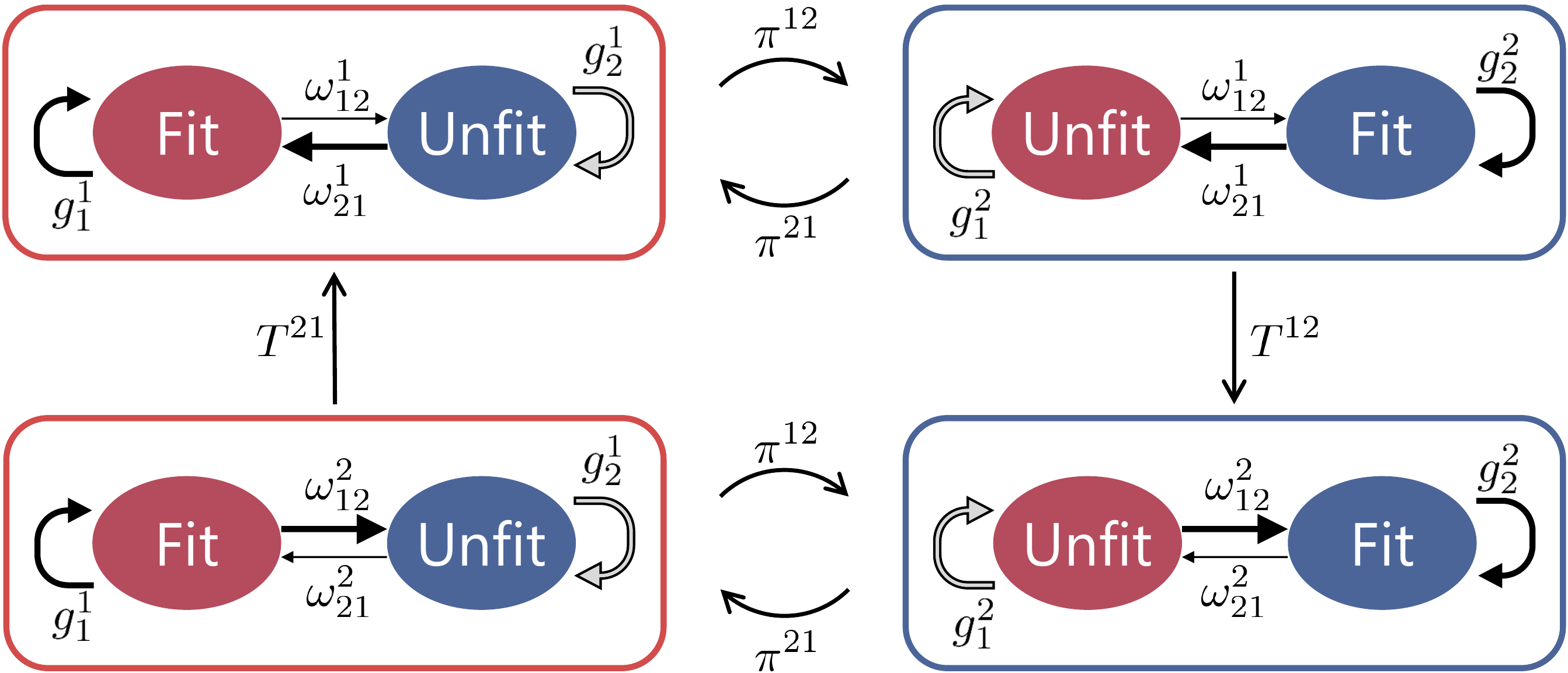} \put(-245,
109){$(\epsilon, \sigma) = (1,1)$} \put(-104, 109){$(2,1)$} \put(-245,
45){$(1,2)$} \put(-104, 45){$(2,2)$} 
\caption{State-transition diagram for delayed adaptation under
environmental fluctuations. Suppose that the environment is of
the first type and the population's knowledge is currently updated, i.e.,
$\epsilon = \sigma = 1$ (the upper-left regime in the figure). Once the
environment~$\epsilon$ changes from~$1$ to~$2$ (the upper-right regime), it
takes $T^{12}$ units of time for the population's knowledge $\sigma$ to be
updated to $2$, which results in the transition to the lower-right regime.} \label{fig:delayedAdaptation}
\end{figure}
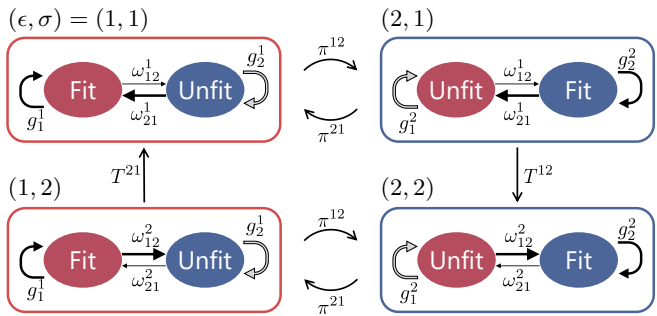

In what follows, we propose a model in which the population senses changes in the environment after a stochastic delay. In particular, at a given time, the environment may change from state $i$ to state $j$, but this change will not be sensed by the colony during a stochastic period of time (i.e., a sensing delay) during which the dynamical behavior of the colony will not adapt to the new environment. To model this behavior, we will denote the knowledge (or belief) of the colony about the environment by $\sigma(t)$, i.e., at time $t$ the colony believes the environment is at state $\sigma(t)$, which may differ from the real state of the environment. In what follows, we describe the random process proposed to model the stochastic sensing delay (see Fig.~\ref{fig:delayedAdaptation} for a schematic picture):
\begin{enumerate}
\item When the environment $\epsilon$ changes from state $i$ to $j$ at time~$t$ (such that
$\sigma(t) = i \neq j$), a random number $T^{ij}$, called the
\emph{response delay}, is drawn from a distribution $X^{ij}$.

\item If the environment remains to be $j$ until time
$t+T^{ij}$, then the population's knowledge~$\sigma$ is updated to
$j$ at time $t+T^{ij}$.

\item However, if the environment $\epsilon$ changes again its state during the period
$(t,t+T^{ij})$, then we discard the random number $T^{ij}$ and go back to the first step.
\end{enumerate}
In our model, we assume \cite{Boulineau2013} that the population's
belief about the environment (represented by the variable~$\sigma$) affects its rate of phenotypic adaptation, while
its growth rates depend on the actual environments and the individuals'
phenotype. Therefore, building on the delay-free
model~\eqref{eq:baseModel}, we propose the following model for the dynamics of the population sizes~$x_k$:
\begin{equation}\label{eq:adaptationDelay}
\frac{dx_k}{dt}
=
g_k^{\epsilon(t)} x_k(t)
+
\sum_{\ell=1}^n\omega_{\ell k}^{\sigma(t)}x_{\ell}(t). 
\end{equation}

For simplicity in our exposition, we focus on the case where there are only two environmental and phenotypic
types ($n=2$), although the analysis presented below can be easily
extended to the general case. Notice that, under this assumption, the
differential equation~\eqref{eq:adaptationDelay} admits the following
vectorial representation
\begin{equation}\label{eq:adaptDelay:MJLS}
\frac{dx}{dt} = A^{(\epsilon(t), \sigma(t))}x(t),
\end{equation}
where the vector~$x$ is defined in \eqref{eq:defx} and, for each
environment-knowledge pair $(\epsilon ,\sigma) = (i, j)$, the matrix
$A^{(i, j)}$ is given by
\begin{equation*}
A^{(i, j)} = \begin{bmatrix}
g^i_1 - \omega^j_{12} & \omega^j_{21}
\\
\omega^j_{12} & g^i_2 - \omega^j_{21}
\end{bmatrix}. 
\end{equation*}

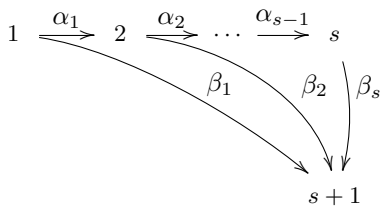
\begin{figure}[tb]
\centering
\newcommand{\myrule}{\rule[-.27cm]{0cm}{.7cm}}%
\centering
\centerline{\xymatrix@!C@=1.9em{
*+={\myrule 1} 
\ar[r]^-{\text{\normalsize$\alpha_1$}}
 \ar@(r,lu)[ddrrr]^-(.6){\text{\normalsize$\beta_1$}}
& 
*+={\myrule 2} 
\ar[r]^-{\text{\normalsize$\alpha_2$}} 
\ar@(r,u)[ddrr]^-(.665){\text{\normalsize$\beta_2$}}
& 
\ \cdots\ 
\ar[r]^-{\text{\normalsize$\alpha_{s-1}$}} 
& 
*+={\myrule s} 
\ar@/^.2cm/[dd]^-(.255){\text{\normalsize$\beta_s$}}
\\
\\
&
&
&
*+={\myrule s+1}
}}
\caption{State transition diagram of the Coxian distribution $C(\alpha, \beta)$. The starting and absorbing states are $1$ and~$s+1$, respectively.}
\label{fig:Coxian}
\end{figure}

In this paper, we allow the response delays $T^{ij}$ to follow a general class
of distributions called Coxian distributions~\cite{Asmussen1996}, defined as
follows: Given two nonnegative vectors $\alpha = (\alpha_1, \dotsc,
\alpha_{s-1})$ and~$\beta = (\beta_1, \dotsc, \beta_s)$, consider the
continuous-time Markov process described by the transition diagram in
Fig.~\ref{fig:Coxian}. The random variable corresponding to the absorption time
of this Markov process into state~$s+1$ starting from state~$1$ follows a Coxian
distribution with parameters $\alpha$ and $\beta$ (see, e.g.,
\cite{Asmussen1996}), denoted by $C(\alpha,\beta)$. It is known that the set of
Coxian distributions is dense in the set of positive-valued
distributions~\cite{Cox1955}. Moreover, there are efficient fitting algorithms
to approximate a given arbitrary distribution by a Coxian
distribution~\cite{Asmussen1996}.

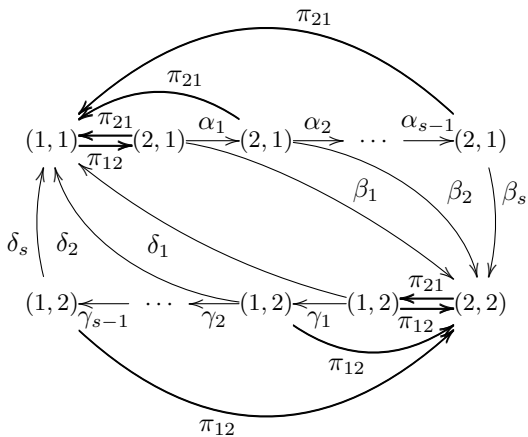
\begin{figure}[tb]
\centering
\newcommand{\myrule}{\rule[-.27cm]{0cm}{.7cm}}%
\centerline{\xymatrix@!C@=1.9em{
*+={\myrule (1,1)} 
\ar@<-.5ex>[r]_-{\text{\normalsize$\pi_{12}$}}
\ar@<-.45ex>[r]
\ar@<-.55ex>[r]
&
*+={\myrule (2,1)} 
\ar[r]^-{\text{\normalsize$\alpha_1$}}
\ar@<-.5ex>[l]_(.4){\text{\normalsize$\pi_{21}$}} 
\ar@<-.45ex>[l] 
\ar@<-.55ex>[l] 
\ar@(r,lu)[ddrrr]^-(.6){\text{\normalsize$\beta_1$}}
& 
*+={\myrule (2,1)} 
\ar[r]^-{\text{\normalsize$\alpha_2$}} 
\ar@/^-.65cm/[ll]_(.4){\text{\normalsize$\pi_{21}$}}
\ar@<.05ex>@/^-.65cm/[ll]
\ar@<-.05ex>@/^-.65cm/[ll]
\ar@(r,u)[ddrr]^-(.665){\text{\normalsize$\beta_2$}}
& 
\ \cdots\ 
\ar[r]^-{\text{\normalsize$\alpha_{s-1}$}} 
& 
*+={\myrule (2,1)} 
\ar@/^.2cm/[dd]^-(.255){\text{\normalsize$\beta_s$}}
\ar@/^-1.5cm/[llll]_(.4){\text{\normalsize$\pi_{21}$}}
\ar@<.05ex>@/^-1.5cm/[llll]
\ar@<-.05ex>@/^-1.5cm/[llll]
\\
\\
*+={\myrule (1,2)} 
\ar@/^.2cm/[uu]^-(.315){\text{\normalsize$\delta_s$}}
\ar@/^-1.5cm/[rrrr]_(.4){\text{\normalsize$\pi_{12}$}}
\ar@<-.05ex>@/^-1.5cm/[rrrr]
\ar@<.05ex>@/^-1.5cm/[rrrr]
&
\ \cdots\ 
\ar[l]^-{\text{\normalsize$\gamma_{s-1}$}}
&
*+={\myrule (1,2)} 
\ar[l]^-{\text{\normalsize$\gamma_2$}} 
\ar@(l,d)[lluu]^-(.7){\text{\normalsize$\delta_2$}}
\ar@/^-.65cm/[rr]_(.4){\text{\normalsize$\pi_{12}$}}
\ar@<.05ex>@/^-.65cm/[rr]
\ar@<-.05ex>@/^-.65cm/[rr]
&
*+={\myrule (1,2)} 
\ar[l]^-{\text{\normalsize$\gamma_1$}} 
\ar@/^.75pc/[uulll]^(.6){\text{\normalsize$\delta_1$}}
\ar@<-.5ex>[r]_(.4){\text{\normalsize$\pi_{12}$}}
\ar@<-.45ex>[r]
\ar@<-.55ex>[r]
&
*+={\myrule (2,2)}
\ar@<-.5ex>[l]_-{\text{\normalsize$\pi_{21}$}}
\ar@<-.55ex>[l]
\ar@<-.45ex>[l]
}}
\caption{Markov chain for the dynamics of the environment-knowledge
pair $(\epsilon, \sigma)$. The thin arrows represent the dynamics of
phase-type distributions, while the thick arrows represent changes in
the environment.} 
\label{fig:directProdMarkov}
\end{figure}

\begin{figure*}[tb]
\centering
\includegraphics[height=\myheight]{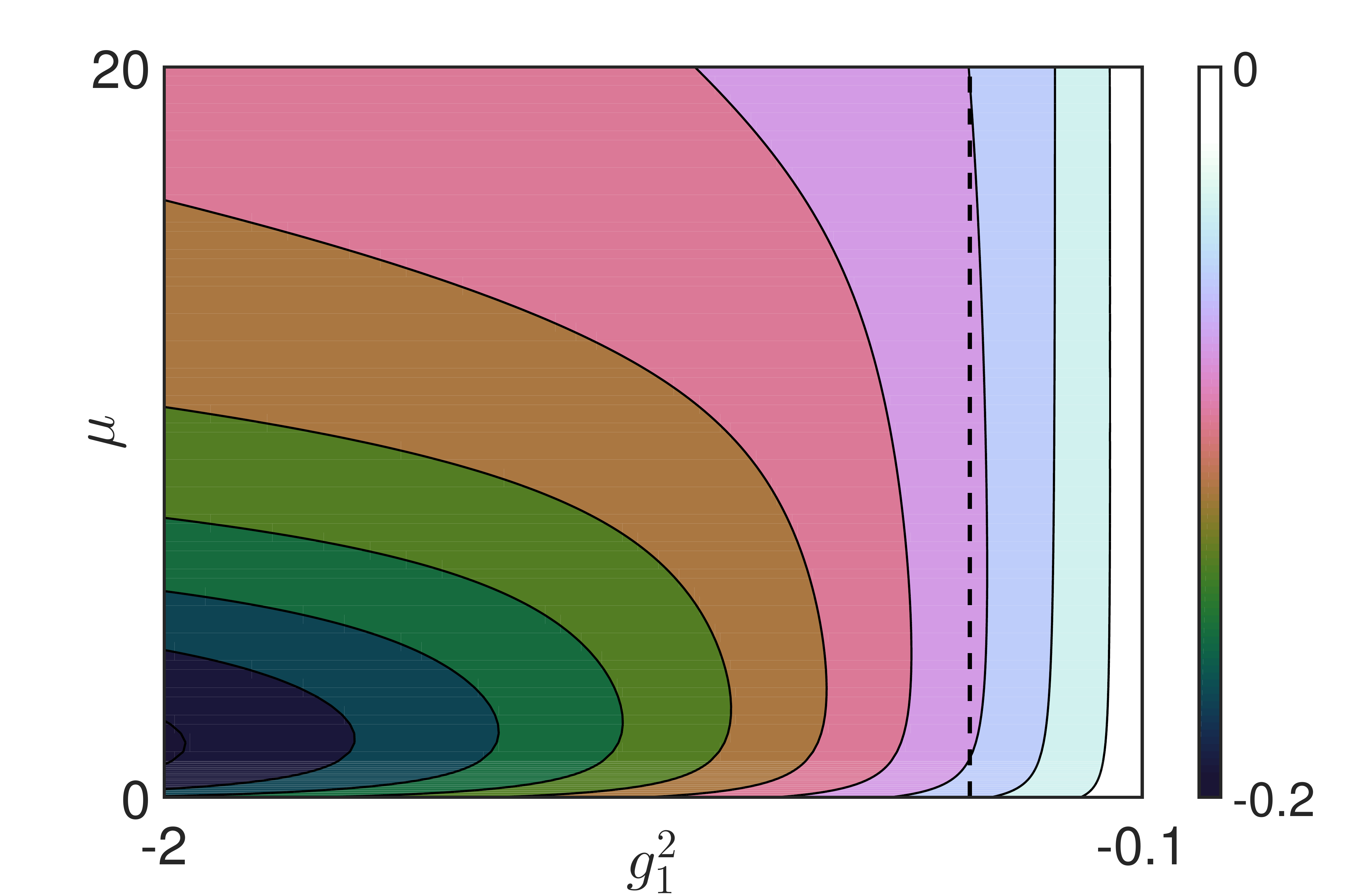} 
\myput{(a)} 
\myspacebetweenfigures 
\includegraphics[height=\myheight]{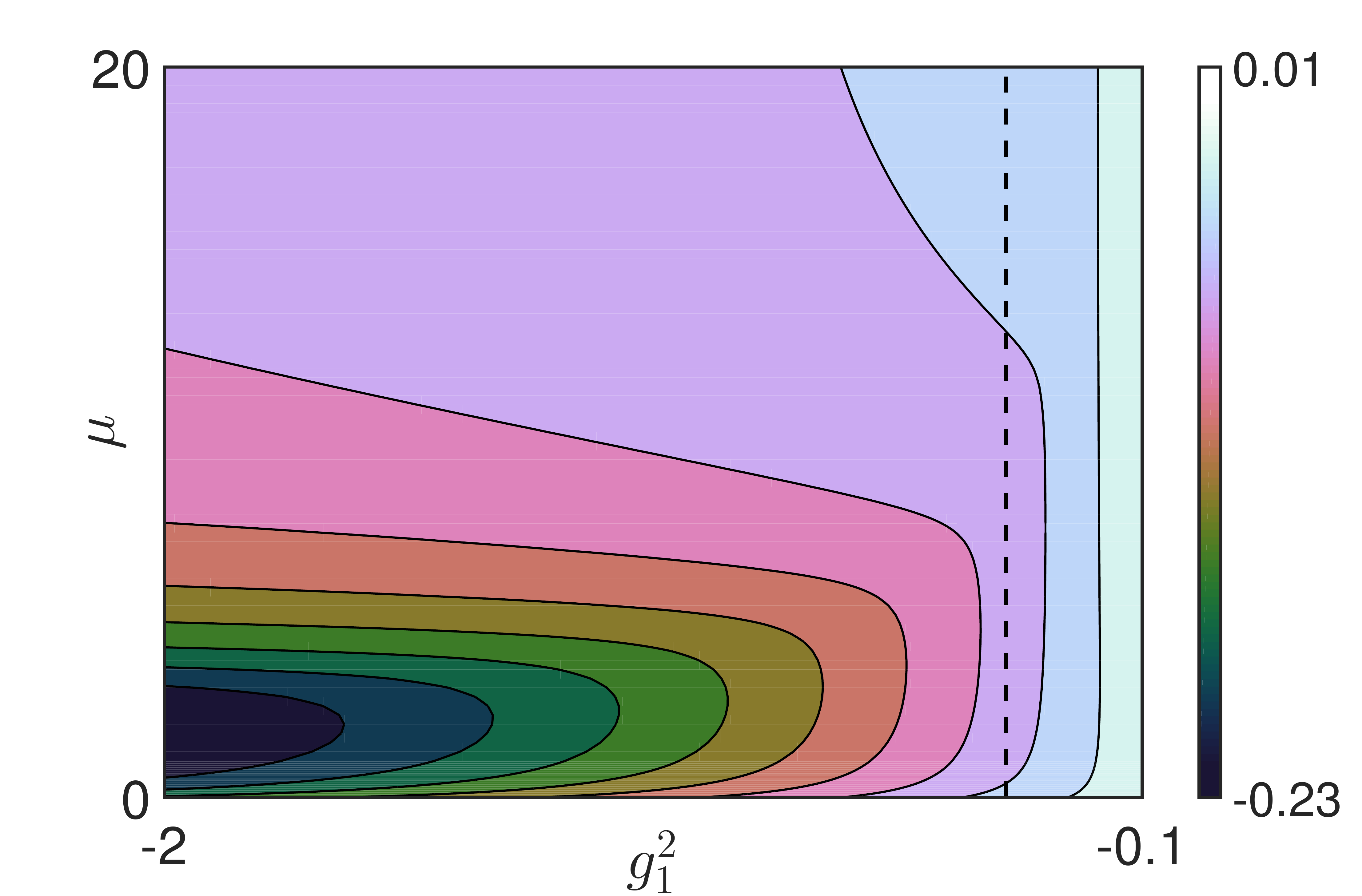} 
\myput{(b)} 
\myspacebetweenfigures 
\includegraphics[height=\myheight]{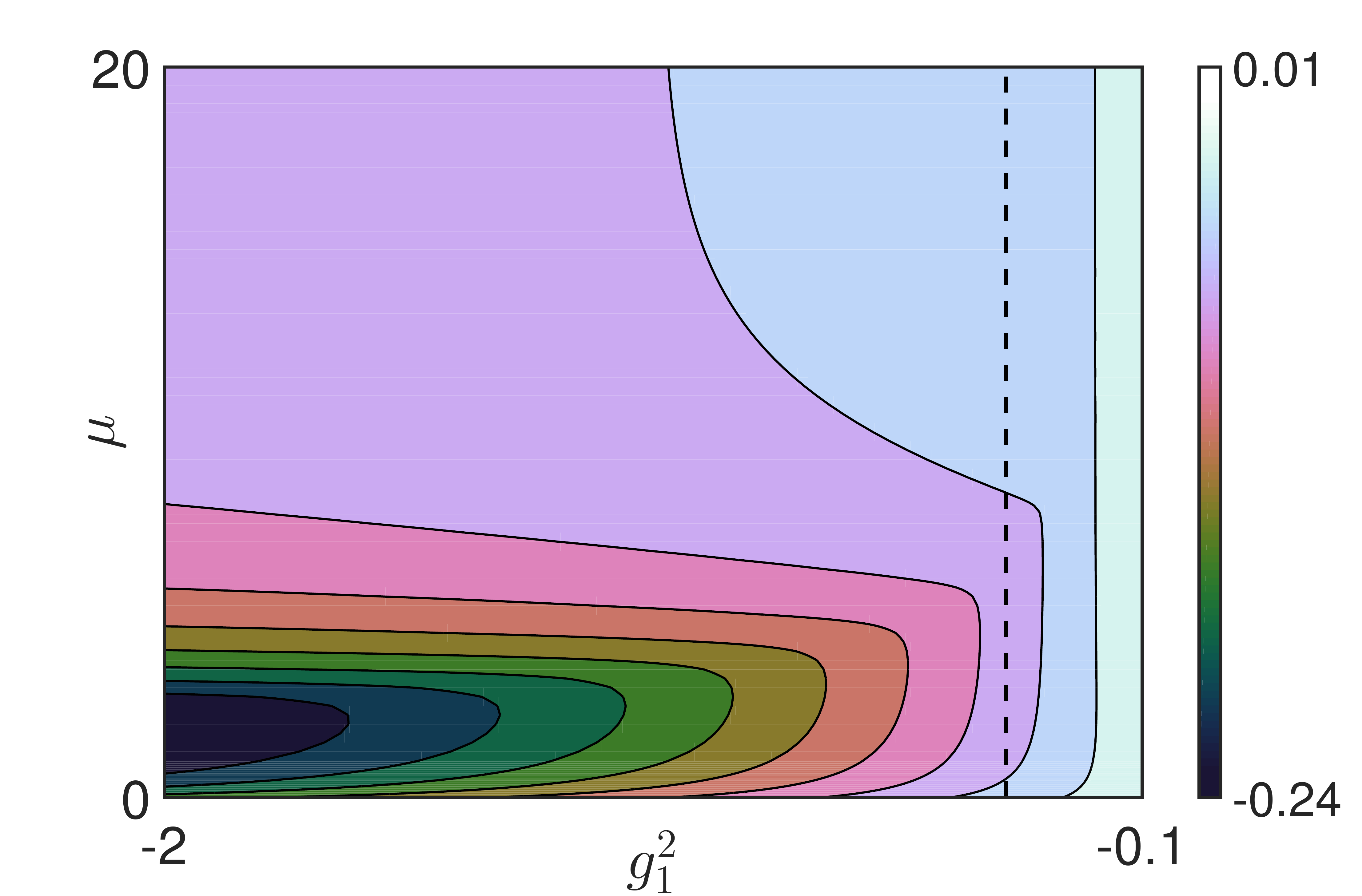} 
\myput{(c)} 
\caption{Growth rates versus $g_1^2$ and~$\mu (=E[T^{12}]=E[T^{21}])$ when $g_2^1=0$: (a) $k = 1$, (b) $k=4$, (c) $k = 16$. Dashed lines indicate the values of $g_1^2$ at which the optimal value of $\mu$ changes. $\mu=0$ is the optimal on the right of the dashed lines, while letting $\mu$ as large as possible yields the optimal growth rates on the left of the dashed lines.} 
\label{fig:adp:mu_b_k}
\end{figure*}

\begin{figure*}[tb]
\centering
\includegraphics[height=\myheight]{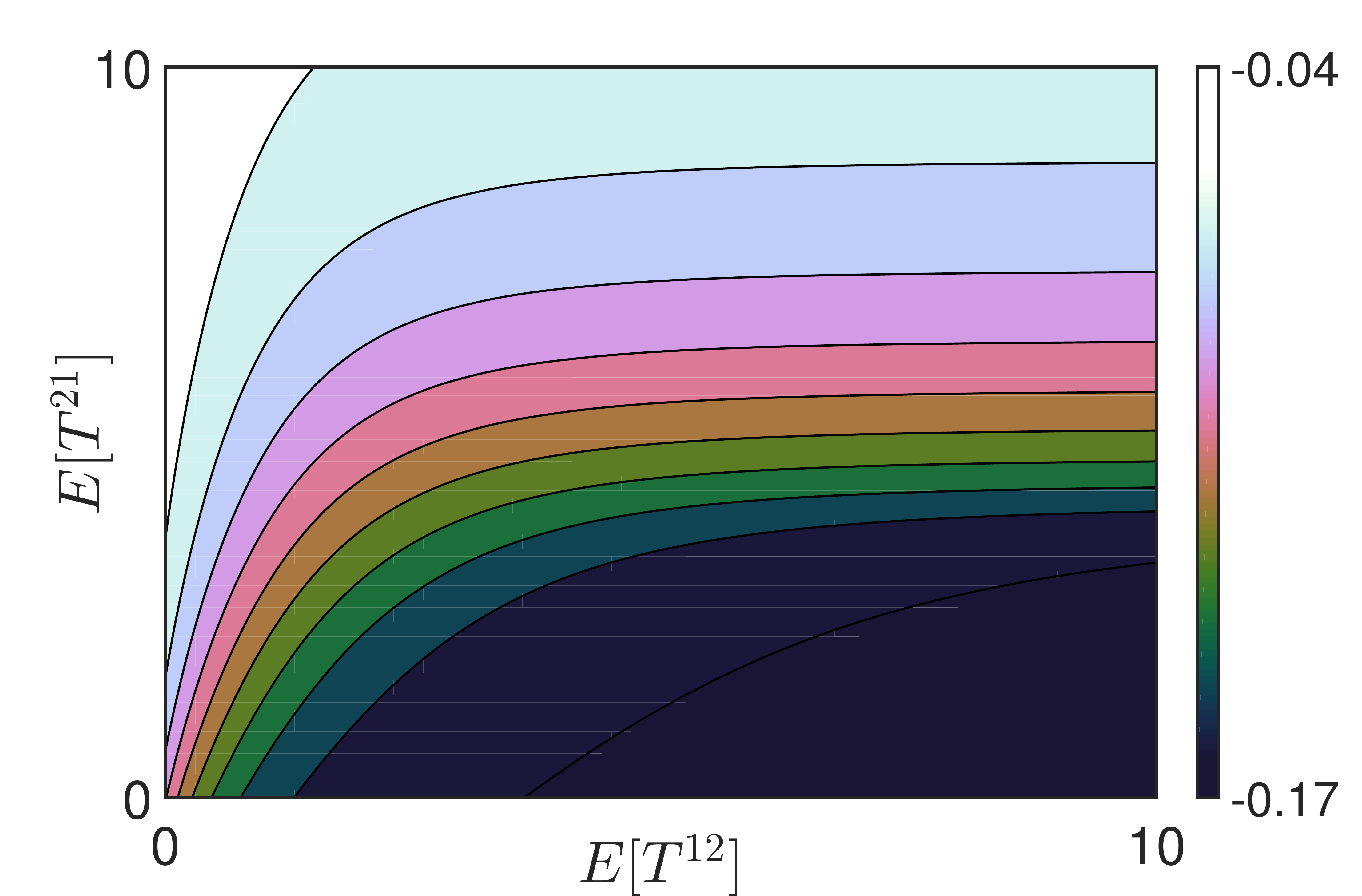}
\myput{(a)}
\myspacebetweenfigures
\includegraphics[height=\myheight]{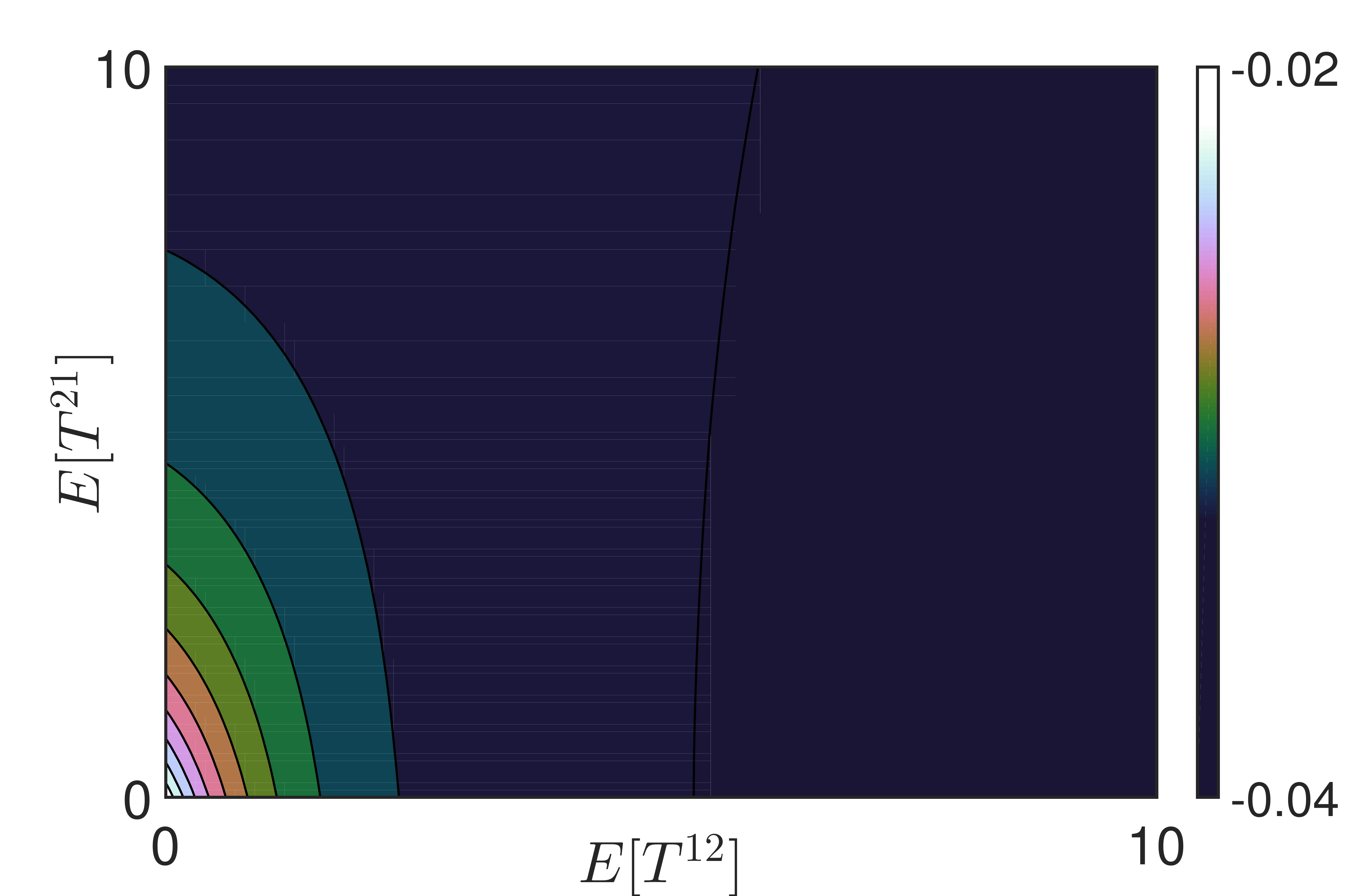}
\myput{(b)}
\myspacebetweenfigures
\includegraphics[height=\myheight]{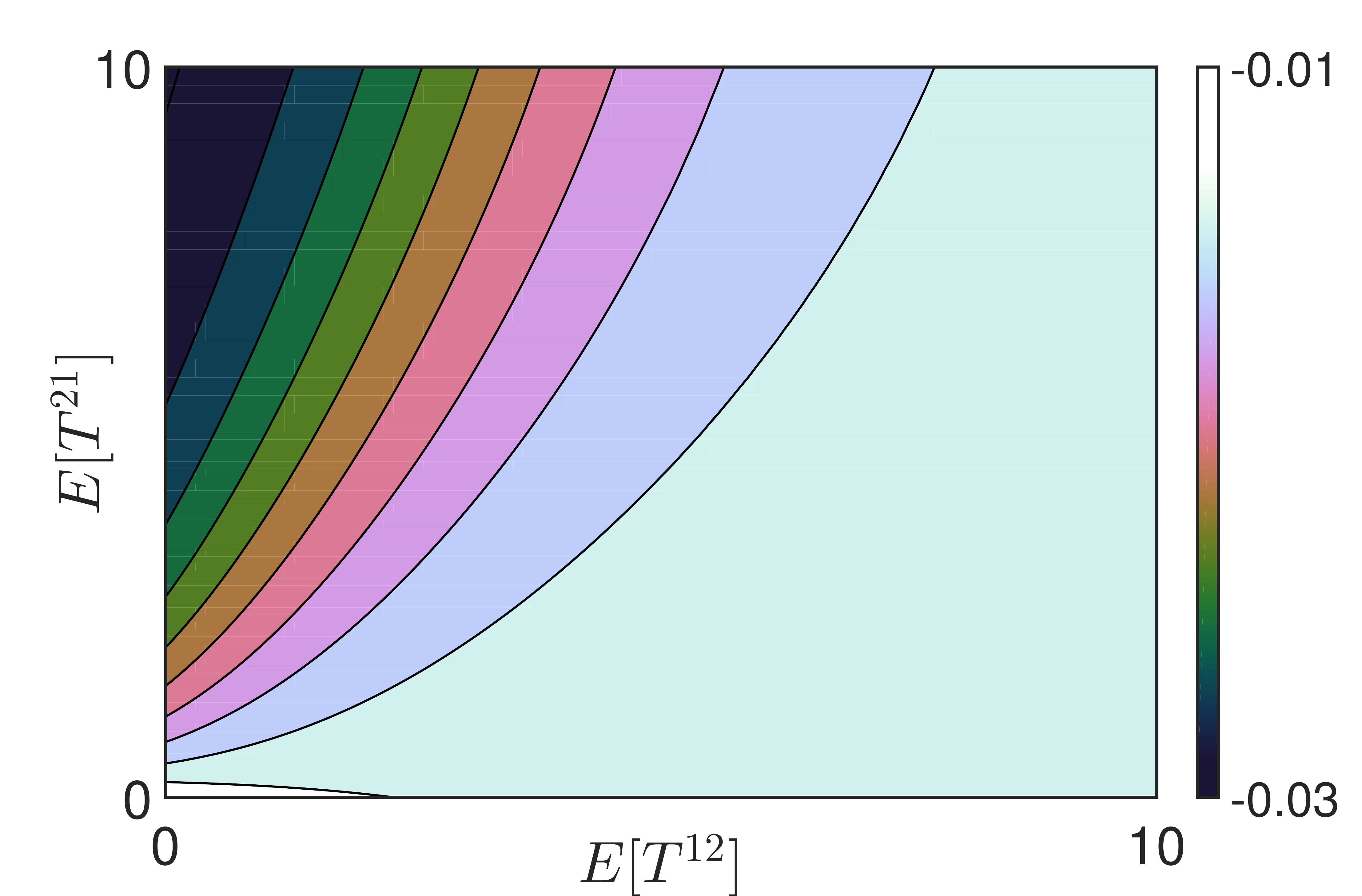}
\myput{(c)}
\caption{Growth rates versus $E[T^{12}]$ and~$E[T^{21}]$ when $g^1_2=0$ and~$k=4$: (a) $g_1^2 = -1$, (b) $g_1^2 = -0.23$, (c) $g_1^2 = -0.15$.}
\label{fig:adp:mu1_mu2}
\end{figure*}

We assume that there exist parameters $\alpha$, $\beta$, as well as two extra vectors $\gamma =
(\gamma_1, \dotsc, \gamma_{s-1})$ and~$\delta = (\delta_1, \dotsc,
\delta_s)$, such that the response delays $T^{12}$ and~$T^{21}$ follow the Coxian distributions $C(\alpha, \beta)$ and~$C(\gamma, \delta)$,
respectively. Then, combining the Markovian dynamics of the
environment~$\epsilon$ with the state transition diagrams corresponding to the the response delays (as seen in Fig.~\ref{fig:Coxian}), we conclude that the dynamics of the environment-knowledge pair~$(\epsilon, \sigma)$ is a
Markov process having the state transition diagram in
Fig.~\ref{fig:directProdMarkov}. This fact implies that
the vectorial population dynamics~\eqref{eq:adaptDelay:MJLS} is a
positive Markov jump linear system~\cite{Ogura2013f}, whose growth
rates can be quantified easily. For this purpose, let us order the
states of the Markov process~$(\epsilon, \sigma)$ as
$\{(1,1),\,(2,1),\,\dotsc,\,(2,1),\, (2,2),\, (1,2),\, \dotsc,\,
(1,2)\}$, where there are $s$ copies of the pairs $(1,2)$ and~$(2,1)$. We then let $\Xi$ be the $(2s+2)\times (2s+2)$
infinitesimal generator matrix of the Markov process~$(\epsilon,
\sigma)$ under this ordering. We finally introduce the matrices $\bar
A^{1} = A^{(1,1)}$, $\bar A^2 = \cdots = \bar A^{s+1} = A^{(2,1)}$,
$\bar A^{s+2} = A^{(2,2)}$, and~$\bar A^{s+3} = \cdots = \bar A^{2s+2}
= A^{(1,2)}$. Then, by a standard result on positive Markov jump
linear systems~\cite[Theorem~5.2]{Ogura2013f}, we can show that the
growth rate of the total population is equal to the maximum real
eigenvalue of the Metzler matrix
\begin{equation*}
\Xi^\top \otimes I_2+ \bigoplus (\bar A^1, \dotsc, \bar A^{2s+2}). 
\end{equation*}
We remark that, by the transition diagram in
Fig.~\ref{fig:directProdMarkov}, the infinitesimal generator~$\Xi$
takes the form
\begin{equation*}
\Xi  = 
\begin{bmatrix}
-\pi^{12}&\pi^{12}u_1^\top&0&O_{1,s}
\\
\pi^{21}\onev&\Xi_\alpha-\pi^{21}I_s-B&\beta&O_{s,s}
\\
O_{s,1}&O_{s,s}&-\pi^{21}&\pi^{21}u_1^\top
\\
\delta&O_{1,s}&\pi^{12}\onev& \Xi_\gamma-D-\pi^{12}I_s
\end{bmatrix}, 
\end{equation*}
where the symbol~$\onev$ denotes the $s$-dimensional column vector of all ones, the matrices~$\Xi_\alpha$ and~$\Xi_\gamma$ are defined by the
formula
\begin{equation*}
\begin{aligned}
\Xi_v &= \begin{bmatrix}
O_{s-1,1}&V
\\
0&O_{1,s-1}
\end{bmatrix} - \ \begin{bmatrix}
V&O_{s-1,1}
\\
O_{1,s-1}&0
\end{bmatrix},
\\
V&=\bigoplus(v_1, \dotsc, v_{s-1}), 
\end{aligned}
\end{equation*}
and~$B, D$ are the diagonal matrices given by $B = \bigoplus(\beta_1, \dotsc, \beta_s)$ and~$D = \bigoplus(\delta_1, \dotsc, \delta_s)$.

\subsection{Numerical simulations}

In this section, we illustrate our results via numerical simulations. We fix the following parameters of the
delay-free model~\eqref{eq:baseModel}: $g_1^1=0.05$ and $g_2^2 =
-0.1$, while the growth rates~$g_2^1$ and~$g_1^2$ are considered free variables. We assume that the response delays $T^{12}$ and~$T^{21}$
follow Erlang distributions with shape~$k$ and mean~$E[T^{12}]$
($E[T^{21}]$, respectively). These distributions are the $k$-sum of
independent and identical exponential distributions and, therefore,
approximate Gaussian distributions when $k$ is large. We also
assume that the response delay~$T^{12}$ (respectively, $T^{21}$) follows a Coxian distribution with parameters $s=k$, $\alpha_1 = \cdots =
\alpha_{k-1} = \beta_k = \lambda = k/E[T^{12}]$ ($=k/E[T^{12}]$,
respectively), and~$\beta_1 = \cdots = \beta_{k-1} = 0$.

In order to numerically analyze the dependency of the growth rates on the relevant
parameters, we temporarily let $E[T^{12}] = E[T^{21}] = \mu$ for a
real parameter~$\mu$ and also fix $g_2^1=0$. We then compute the
growth rates of the total population for various values of $\mu$,
$g_1^2$, and~$k$. We show the obtained growth rates in
Fig.~\ref{fig:adp:mu_b_k}. As in the previous section, we observe a non-trivial dependency
of the growth rates on the model parameters. Specifically, we observe that, for all values of $k$, shorter delays improve the growth
rates only when $g_1^2$ is relatively large; while longer delays
increase the overall fitness in the region of small~$g_1^2$. This
phase shift (indicated by dashed lines in the figures) occurs at $g_1^2=-0.44$ ($k=1$) and~$g_1^2=-0.37$ ($k=4$ and~$k =
16$), suggesting that an immediate sensing of environmental
changes does not necessarily improve the overall fitness of the
population.

We then allow the means $E[T^{12}]$ and $E[T^{21}]$ of the response
delays to be different and examine the dependency of the growth rates
on these two delays for different values of $g_1^2$, while fixing
$k=4$ and~$g_2^1=0$. The computed growth rates are shown in
Fig.~\ref{fig:adp:mu1_mu2}. As seen in Fig.~\ref{fig:adp:mu_b_k}, we
observe that shorter adaptation
delays do not necessary improve the overall fitness of the population.
In contrast, for the case of $g_1^2 = -1$ (Fig.~\ref{fig:adp:mu1_mu2}(a)),
the growth rate increases as $E[T^{12}]$ decreases (although this does not
happen for the other delay~$E[T^{21}]$). On the other hand, we can
observe the opposite phenomena in the case of $g_1^2 = -0.15$
(Fig.~\ref{fig:adp:mu1_mu2}(c)). Finally, for the case of $g_1^2 = -0.23$
(Fig.~\ref{fig:adp:mu1_mu2}(b)), the growth rate increases when
both expected delays decrease. These simulations illustrate
that the optimal length of the adaptation delays (from the perspective of
maximizing the growth rate) depends, in a
non-trivial and sometimes counterintuitive way, on the various parameters of the model.

\begin{figure}[tb]
\centering
\includegraphics[width=.9\linewidth]{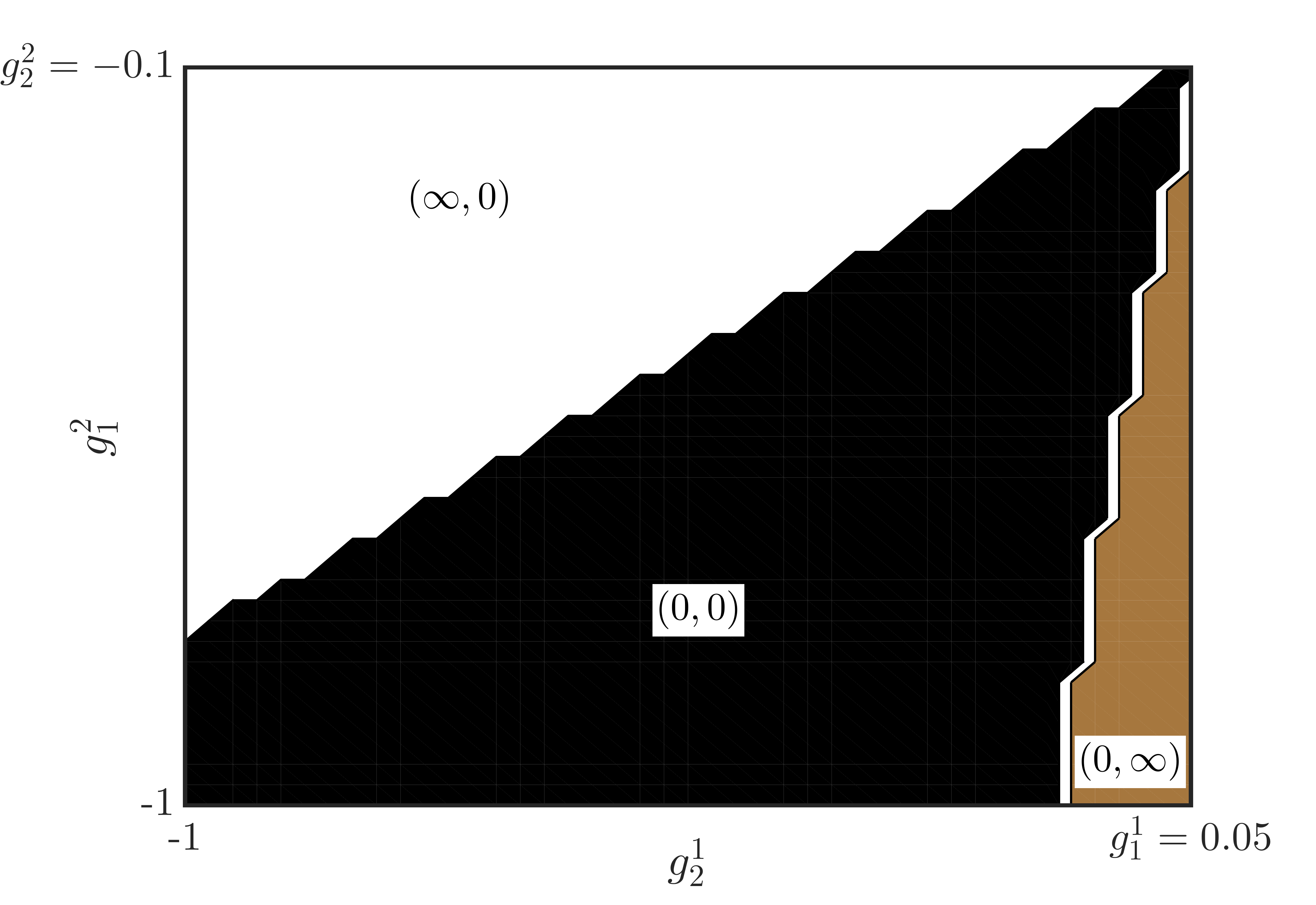}
\hspace{.35cm}
\caption{Phase diagram of the optimal length of expected adaptation delays $(E[T^{12}], E[T^{21}])$ versus $g_2^1$ and~$g_1^2$.}
\label{fig:phaseDiagram}
\end{figure}

In order to understand how the optimality of expected delays are
affected by other parameters, we calculate a phase diagram for
delay optimization. Fig.~\ref{fig:phaseDiagram} presents a phase
diagram showing how the optimal length of the pair of expected delays
$(E[T^{12}], E[T^{21}])$ vary depending on the pair~$(g_2^1, g_1^2)$.
We observe that, in the white region (where $g_1^2$ is relatively
large), it is optimal for the population to immediately adapt to the
environment $1$ but never recognize the change to the environment $2$.
This phenomena happens because a very large $g_1^2$ makes the fitness
of the population in the environment-knowledge pair~$(2, 1)$ (the
upper-right regime in Fig.~\ref{fig:delayedAdaptation}) the largest
among others, making it optimal for the pair \emph{not} to transit
from $(2, 1)$ to $(2,2)$. We can observe the same phenomena in the
lower-right region in the figure (where $g_2^1$ is relatively large). We
can also see that, in the the black region (in
which the values of $g_1^2$ and $g_2^1$ are ``balanced''), it is optimal to
immediately adapt to environmental changes, as one would
expect.

\section{Conclusion}  \label{sec:Conclusion}

In this paper, we have proposed an analytical framework to precisely quantify
the fitness of bet-hedging populations, such as bacterial colonies, whose
reaction to environmental fluctuations exhibit time-delays. We have specifically
considered the situation where delays are present in either the proliferation or
the adaptation of the individuals. For both cases, we have developed efficient
techniques to quantify the growth rates of bet-hedging populations using the
maximum real eigenvalues of certain Metzler matrices. We have confirmed the
accuracy of the proposed methods by numerical simulations and show that, in
harsh environments, it is better for individuals to have delays in their
proliferations; while in favorable environments, immediate reproductions would
be a better strategy to increase the overall fitness of the population. These
simulations also illustrate how the growth rates of bet-hedging populations with
delays depend on relevant parameters in a highly non-trivial, sometimes
counterintuitive, manner. In particular, we have found that, in certain
situations, fast sensing of environmental changes or shorter adaptation delays
do not necessarily improve the overall fitness of the population.

Although we have assumed in this paper that the environment
dynamically switches according to a time-homogeneous Markov process, it is more
realistic to consider environmental changes driven by non-Markovian stochastic
processes such as those having
pseudo-periodicity~\cite{Chichigina2011,Kargovsky2015}. A possible direction for
future research is to consider these non-Markovian transitions between
environments, and examine how the non-Markovian properties alter fitness
landscapes.

\section*{Acknowledgement}

This work was supported in part by the NSF under Grants CAREER-ECCS-1651433 and IIS-1447470.

\appendix*

\section{}

From the stochastic differential equation~\eqref{eq:d(del.ox.x).pre},
we immediately see that the expectation $\zeta = E[z]$ obeys the following
differential equation
\begin{equation}\label{eq:dbarx/dt}
\begin{aligned}
&\frac{d\zeta}{dt} = 
\sum_{i=1}^nE[(\eta\otimes I_n)\eta^i(\mathcal{A}^i x)] 
\\
&\myindent\myindent
+ \sum_{i=1}^n \sum_{j\neq i}E\left[\bigl((U_{ji}-U_{ii})\eta\bigr)\otimes x \right]\pi^{ij}.
\end{aligned}
\end{equation}
It is shown in the proof of \cite[Proposition~5.3]{Ogura2013f} that
the second term of the right-hand side of \eqref{eq:dbarx/dt} equals
$(\Pi^\top \otimes I_n)\zeta$. Let us compute the first term. Since
$\eta^i \eta^j =0$ for $i\neq j$ and~$(\eta^i)^2=\eta^i$ by the
definition of $\eta$, we have $\eta^i \eta = \eta^i u_i$. Therefore,
it follows that $(\eta \otimes I_n) \eta^i({\mathcal{A}}^i x) = (u_i
\otimes A^i)\eta^i x + (u_i\otimes I_n) \eta^i(B^ix)$. Hence, we can
compute the first term in the right-hand side of \eqref{eq:dbarx/dt}
as
\begin{equation}\label{eq:dbarx/dt2}
\begin{aligned}
&\sum_{i=1}^n E[(\eta\otimes I_n)\eta^i(\mathcal{A}^i x)]
\\
&\myindent\myindent
=\biggl(\bigoplus_{i=1}^n A^i\biggr) \zeta
+
\sum_{i=1}^n(u_i\otimes I)E[\eta^i(B^ix)].
\end{aligned}
\end{equation}
 We need to evaluate the second term of the right-hand side of this
 equation. We notice that a simple application of the tower property
 on conditional expectations (see, e.g., \cite{Borkar1995}) shows
 $E[\eta^i(t)x(t-h)]=((u_i^\top e^{\Pi^\top h})\otimes I_n)
 \zeta(t-h)$. We can therefore prove that $E[\eta^i(t)(B^ix)(t)] =
 \sum_{k=1}^n (p^i_k (U_{ki} e^{\Pi^\top d^i_k})\otimes
 u_k^\top)\zeta(t-d^i_k)$. This argument indeed shows that the second term
 in the right-hand side of \eqref{eq:dbarx/dt2} equals $\sum_{i,k=1}^n
 \mathbf A^i_k\zeta(t-d^i_k)$, completing the proof of the
 differential equation~\eqref{eq:barMJLSD}.


\begin{thebibliography}{26}%
\makeatletter
\providecommand \@ifxundefined [1]{%
 \@ifx{#1\undefined}
}%
\providecommand \@ifnum [1]{%
 \ifnum #1\expandafter \@firstoftwo
 \else \expandafter \@secondoftwo
 \fi
}%
\providecommand \@ifx [1]{%
 \ifx #1\expandafter \@firstoftwo
 \else \expandafter \@secondoftwo
 \fi
}%
\providecommand \natexlab [1]{#1}%
\providecommand \enquote  [1]{``#1''}%
\providecommand \bibnamefont  [1]{#1}%
\providecommand \bibfnamefont [1]{#1}%
\providecommand \citenamefont [1]{#1}%
\providecommand \href@noop [0]{\@secondoftwo}%
\providecommand \href [0]{\begingroup \@sanitize@url \@href}%
\providecommand \@href[1]{\@@startlink{#1}\@@href}%
\providecommand \@@href[1]{\endgroup#1\@@endlink}%
\providecommand \@sanitize@url [0]{\catcode `\\12\catcode `\$12\catcode
  `\&12\catcode `\#12\catcode `\^12\catcode `\_12\catcode `\%12\relax}%
\providecommand \@@startlink[1]{}%
\providecommand \@@endlink[0]{}%
\providecommand \url  [0]{\begingroup\@sanitize@url \@url }%
\providecommand \@url [1]{\endgroup\@href {#1}{\urlprefix }}%
\providecommand \urlprefix  [0]{URL }%
\providecommand \Eprint [0]{\href }%
\providecommand \doibase [0]{http://dx.doi.org/}%
\providecommand \selectlanguage [0]{\@gobble}%
\providecommand \bibinfo  [0]{\@secondoftwo}%
\providecommand \bibfield  [0]{\@secondoftwo}%
\providecommand \translation [1]{[#1]}%
\providecommand \BibitemOpen [0]{}%
\providecommand \bibitemStop [0]{}%
\providecommand \bibitemNoStop [0]{.\EOS\space}%
\providecommand \EOS [0]{\spacefactor3000\relax}%
\providecommand \BibitemShut  [1]{\csname bibitem#1\endcsname}%
\let\auto@bib@innerbib\@empty
\bibitem [{\citenamefont {Kussell}\ and\ \citenamefont
  {Leibler}(2005)}]{Kussell2005}%
  \BibitemOpen
  \bibfield  {author} {\bibinfo {author} {\bibfnamefont {E.}~\bibnamefont
  {Kussell}}\ and\ \bibinfo {author} {\bibfnamefont {S.}~\bibnamefont
  {Leibler}},\ }\href {\doibase 10.1126/science.1114383} {\bibfield  {journal}
  {\bibinfo  {journal} {Science}\ }\textbf {\bibinfo {volume} {309}},\ \bibinfo
  {pages} {2075} (\bibinfo {year} {2005})}\BibitemShut {NoStop}%
\bibitem [{\citenamefont {Acar}\ \emph {et~al.}(2008)\citenamefont {Acar},
  \citenamefont {Mettetal},\ and\ \citenamefont {van Oudenaarden}}]{Acar2008}%
  \BibitemOpen
  \bibfield  {author} {\bibinfo {author} {\bibfnamefont {M.}~\bibnamefont
  {Acar}}, \bibinfo {author} {\bibfnamefont {J.~T.}\ \bibnamefont {Mettetal}},
  \ and\ \bibinfo {author} {\bibfnamefont {A.}~\bibnamefont {van
  Oudenaarden}},\ }\href {\doibase 10.1038/ng.110} {\bibfield  {journal}
  {\bibinfo  {journal} {Nature Genetics}\ }\textbf {\bibinfo {volume} {40}},\
  \bibinfo {pages} {471} (\bibinfo {year} {2008})}\BibitemShut {NoStop}%
\bibitem [{\citenamefont {Seger}\ and\ \citenamefont
  {Brockmann}(1987)}]{Seger1987}%
  \BibitemOpen
  \bibfield  {author} {\bibinfo {author} {\bibfnamefont {J.}~\bibnamefont
  {Seger}}\ and\ \bibinfo {author} {\bibfnamefont {H.~J.}\ \bibnamefont
  {Brockmann}},\ }in\ \href@noop {} {\emph {\bibinfo {booktitle} {Oxford
  Surveys in Evolutionary Biology}}},\ Vol.~\bibinfo {volume} {4}\ (\bibinfo
  {publisher} {Oxford University Press},\ \bibinfo {year} {1987})\ pp.\
  \bibinfo {pages} {182--211}\BibitemShut {NoStop}%
\bibitem [{\citenamefont {Boulineau}\ \emph {et~al.}(2013)\citenamefont
  {Boulineau}, \citenamefont {Tostevin}, \citenamefont {Kiviet}, \citenamefont
  {ten Wolde}, \citenamefont {Nghe},\ and\ \citenamefont
  {Tans}}]{Boulineau2013}%
  \BibitemOpen
  \bibfield  {author} {\bibinfo {author} {\bibfnamefont {S.}~\bibnamefont
  {Boulineau}}, \bibinfo {author} {\bibfnamefont {F.}~\bibnamefont {Tostevin}},
  \bibinfo {author} {\bibfnamefont {D.~J.}\ \bibnamefont {Kiviet}}, \bibinfo
  {author} {\bibfnamefont {P.~R.}\ \bibnamefont {ten Wolde}}, \bibinfo {author}
  {\bibfnamefont {P.}~\bibnamefont {Nghe}}, \ and\ \bibinfo {author}
  {\bibfnamefont {S.~J.}\ \bibnamefont {Tans}},\ }\href@noop {} {\bibfield
  {journal} {\bibinfo  {journal} {PLoS ONE}\ }\textbf {\bibinfo {volume} {8}}
  (\bibinfo {year} {2013})}\BibitemShut {NoStop}%
\bibitem [{\citenamefont {Oppenheim}\ \emph {et~al.}(2005)\citenamefont
  {Oppenheim}, \citenamefont {Kobiler}, \citenamefont {Stavans}, \citenamefont
  {Court},\ and\ \citenamefont {Adhya}}]{Oppenheim2005}%
  \BibitemOpen
  \bibfield  {author} {\bibinfo {author} {\bibfnamefont {A.~B.}\ \bibnamefont
  {Oppenheim}}, \bibinfo {author} {\bibfnamefont {O.}~\bibnamefont {Kobiler}},
  \bibinfo {author} {\bibfnamefont {J.}~\bibnamefont {Stavans}}, \bibinfo
  {author} {\bibfnamefont {D.~L.}\ \bibnamefont {Court}}, \ and\ \bibinfo
  {author} {\bibfnamefont {S.}~\bibnamefont {Adhya}},\ }\href {\doibase
  10.1146/annurev.genet.39.073003.113656} {\bibfield  {journal} {\bibinfo
  {journal} {Annual Review of Genetics}\ }\textbf {\bibinfo {volume} {39}},\
  \bibinfo {pages} {409} (\bibinfo {year} {2005})}\BibitemShut {NoStop}%
\bibitem [{\citenamefont {Gremer}\ and\ \citenamefont
  {Venable}(2014)}]{Gremer2014}%
  \BibitemOpen
  \bibfield  {author} {\bibinfo {author} {\bibfnamefont {J.~R.}\ \bibnamefont
  {Gremer}}\ and\ \bibinfo {author} {\bibfnamefont {D.~L.}\ \bibnamefont
  {Venable}},\ }\href {\doibase 10.1111/ele.12241} {\bibfield  {journal}
  {\bibinfo  {journal} {Ecology Letters}\ }\textbf {\bibinfo {volume} {17}},\
  \bibinfo {pages} {380} (\bibinfo {year} {2014})}\BibitemShut {NoStop}%
\bibitem [{\citenamefont {van~der Woude}\ and\ \citenamefont
  {Baumler}(2004)}]{Woude2004}%
  \BibitemOpen
  \bibfield  {author} {\bibinfo {author} {\bibfnamefont {M.~W.}\ \bibnamefont
  {van~der Woude}}\ and\ \bibinfo {author} {\bibfnamefont {A.~J.}\ \bibnamefont
  {Baumler}},\ }\href {\doibase 10.1128/CMR.17.3.581-611.2004} {\bibfield
  {journal} {\bibinfo  {journal} {Clinical Microbiology Reviews}\ }\textbf
  {\bibinfo {volume} {17}},\ \bibinfo {pages} {581} (\bibinfo {year}
  {2004})}\BibitemShut {NoStop}%
\bibitem [{\citenamefont {Belete}\ and\ \citenamefont
  {Bal\'azsi}(2015)}]{Belete2015}%
  \BibitemOpen
  \bibfield  {author} {\bibinfo {author} {\bibfnamefont {M.~K.}\ \bibnamefont
  {Belete}}\ and\ \bibinfo {author} {\bibfnamefont {G.}~\bibnamefont
  {Bal\'azsi}},\ }\href {\doibase 10.1103/PhysRevE.92.062716} {\bibfield
  {journal} {\bibinfo  {journal} {Physical Review E}\ }\textbf {\bibinfo
  {volume} {92}},\ \bibinfo {pages} {062716} (\bibinfo {year}
  {2015})}\BibitemShut {NoStop}%
\bibitem [{\citenamefont {Ga\'al}\ \emph {et~al.}(2010)\citenamefont {Ga\'al},
  \citenamefont {Pitchford},\ and\ \citenamefont {Wood}}]{Gaal2010}%
  \BibitemOpen
  \bibfield  {author} {\bibinfo {author} {\bibfnamefont {B.}~\bibnamefont
  {Ga\'al}}, \bibinfo {author} {\bibfnamefont {J.~W.}\ \bibnamefont
  {Pitchford}}, \ and\ \bibinfo {author} {\bibfnamefont {A.~J.}\ \bibnamefont
  {Wood}},\ }\href {\doibase 10.1534/genetics.109.113431} {\bibfield  {journal}
  {\bibinfo  {journal} {Genetics}\ }\textbf {\bibinfo {volume} {184}},\
  \bibinfo {pages} {1113} (\bibinfo {year} {2010})}\BibitemShut {NoStop}%
\bibitem [{\citenamefont {M\"uller}\ \emph {et~al.}(2013)\citenamefont
  {M\"uller}, \citenamefont {Hense}, \citenamefont {Fuchs}, \citenamefont
  {Utz},\ and\ \citenamefont {P{\"{o}}tzsche}}]{Muller2013a}%
  \BibitemOpen
  \bibfield  {author} {\bibinfo {author} {\bibfnamefont {J.}~\bibnamefont
  {M\"uller}}, \bibinfo {author} {\bibfnamefont {B.}~\bibnamefont {Hense}},
  \bibinfo {author} {\bibfnamefont {T.}~\bibnamefont {Fuchs}}, \bibinfo
  {author} {\bibfnamefont {M.}~\bibnamefont {Utz}}, \ and\ \bibinfo {author}
  {\bibfnamefont {C.}~\bibnamefont {P{\"{o}}tzsche}},\ }\href {\doibase
  10.1016/j.jtbi.2013.07.017} {\bibfield  {journal} {\bibinfo  {journal}
  {Journal of Theoretical Biology}\ }\textbf {\bibinfo {volume} {336}},\
  \bibinfo {pages} {144} (\bibinfo {year} {2013})}\BibitemShut {NoStop}%
\bibitem [{\citenamefont {Skanata}\ and\ \citenamefont
  {Kussell}(2016)}]{Skanata2016}%
  \BibitemOpen
  \bibfield  {author} {\bibinfo {author} {\bibfnamefont {A.}~\bibnamefont
  {Skanata}}\ and\ \bibinfo {author} {\bibfnamefont {E.}~\bibnamefont
  {Kussell}},\ }\href {\doibase 10.1103/PhysRevLett.117.038104} {\bibfield
  {journal} {\bibinfo  {journal} {Physical Review Letters}\ }\textbf {\bibinfo
  {volume} {117}},\ \bibinfo {pages} {038104} (\bibinfo {year}
  {2016})}\BibitemShut {NoStop}%
\bibitem [{\citenamefont {Sughiyama}\ and\ \citenamefont
  {Kobayashi}(2017)}]{Sughiyama2017}%
  \BibitemOpen
  \bibfield  {author} {\bibinfo {author} {\bibfnamefont {Y.}~\bibnamefont
  {Sughiyama}}\ and\ \bibinfo {author} {\bibfnamefont {T.~J.}\ \bibnamefont
  {Kobayashi}},\ }\href {\doibase 10.1103/PhysRevE.95.012131} {\bibfield
  {journal} {\bibinfo  {journal} {Physical Review E}\ }\textbf {\bibinfo
  {volume} {95}},\ \bibinfo {pages} {012131} (\bibinfo {year}
  {2017})}\BibitemShut {NoStop}%
\bibitem [{\citenamefont {Stumpf}\ \emph {et~al.}(2002)\citenamefont {Stumpf},
  \citenamefont {Laidlaw},\ and\ \citenamefont {Jansen}}]{Stumpf2002}%
  \BibitemOpen
  \bibfield  {author} {\bibinfo {author} {\bibfnamefont {M.~P.~H.}\
  \bibnamefont {Stumpf}}, \bibinfo {author} {\bibfnamefont {Z.}~\bibnamefont
  {Laidlaw}}, \ and\ \bibinfo {author} {\bibfnamefont {V.~A.~A.}\ \bibnamefont
  {Jansen}},\ }\href {\doibase 10.1073/pnas.232546899} {\bibfield  {journal}
  {\bibinfo  {journal} {Proceedings of the National Academy of Sciences USA}\
  }\textbf {\bibinfo {volume} {99}},\ \bibinfo {pages} {15234} (\bibinfo {year}
  {2002})}\BibitemShut {NoStop}%
\bibitem [{\citenamefont {Baker}\ \emph {et~al.}(1998)\citenamefont {Baker},
  \citenamefont {Bocharov}, \citenamefont {Paul},\ and\ \citenamefont
  {Rihan}}]{Baker1998}%
  \BibitemOpen
  \bibfield  {author} {\bibinfo {author} {\bibfnamefont {C.~T.~H.}\
  \bibnamefont {Baker}}, \bibinfo {author} {\bibfnamefont {G.~A.}\ \bibnamefont
  {Bocharov}}, \bibinfo {author} {\bibfnamefont {C.~A.~H.}\ \bibnamefont
  {Paul}}, \ and\ \bibinfo {author} {\bibfnamefont {F.~A.}\ \bibnamefont
  {Rihan}},\ }\href@noop {} {\bibfield  {journal} {\bibinfo  {journal} {Journal
  of Mathematical Biology}\ }\textbf {\bibinfo {volume} {37}},\ \bibinfo
  {pages} {341} (\bibinfo {year} {1998})}\BibitemShut {NoStop}%
\bibitem [{\citenamefont {Thattai}\ and\ \citenamefont {{Van
  Oudenaarden}}(2004)}]{Thattai2004}%
  \BibitemOpen
  \bibfield  {author} {\bibinfo {author} {\bibfnamefont {M.}~\bibnamefont
  {Thattai}}\ and\ \bibinfo {author} {\bibfnamefont {A.}~\bibnamefont {{Van
  Oudenaarden}}},\ }\href {\doibase 10.1534/genetics.167.1.523} {\bibfield
  {journal} {\bibinfo  {journal} {Genetics}\ }\textbf {\bibinfo {volume}
  {167}},\ \bibinfo {pages} {523} (\bibinfo {year} {2004})}\BibitemShut
  {NoStop}%
\bibitem [{\citenamefont {Brockett}(2009)}]{Brockett2009}%
  \BibitemOpen
  \bibfield  {author} {\bibinfo {author} {\bibfnamefont {R.~W.}\ \bibnamefont
  {Brockett}},\ }\href
  {http://www.eeci-institute.eu/pdf/M015/RogersStochastic.pdf} {\enquote
  {\bibinfo {title} {{Stochastic Control}},}\ } (\bibinfo {year}
  {2009})\BibitemShut {NoStop}%
\bibitem [{\citenamefont {Ciuchi}\ \emph {et~al.}(1993)\citenamefont {Ciuchi},
  \citenamefont {{de Pasquale}},\ and\ \citenamefont {Spagnolo}}]{Ciuchi1993}%
  \BibitemOpen
  \bibfield  {author} {\bibinfo {author} {\bibfnamefont {S.}~\bibnamefont
  {Ciuchi}}, \bibinfo {author} {\bibfnamefont {F.}~\bibnamefont {{de
  Pasquale}}}, \ and\ \bibinfo {author} {\bibfnamefont {B.}~\bibnamefont
  {Spagnolo}},\ }\href {\doibase 10.1103/PhysRevE.47.3915} {\bibfield
  {journal} {\bibinfo  {journal} {Physical Review E}\ }\textbf {\bibinfo
  {volume} {47}},\ \bibinfo {pages} {3915} (\bibinfo {year}
  {1993})}\BibitemShut {NoStop}%
\bibitem [{\citenamefont {Hasty}\ \emph {et~al.}(2000)\citenamefont {Hasty},
  \citenamefont {Pradines}, \citenamefont {Dolnik},\ and\ \citenamefont
  {Collins}}]{Hasty2000}%
  \BibitemOpen
  \bibfield  {author} {\bibinfo {author} {\bibfnamefont {J.}~\bibnamefont
  {Hasty}}, \bibinfo {author} {\bibfnamefont {J.}~\bibnamefont {Pradines}},
  \bibinfo {author} {\bibfnamefont {M.}~\bibnamefont {Dolnik}}, \ and\ \bibinfo
  {author} {\bibfnamefont {J.~J.}\ \bibnamefont {Collins}},\ }\href {\doibase
  10.1073/pnas.040411297} {\bibfield  {journal} {\bibinfo  {journal}
  {Proceedings of the National Academy of Sciences of the United States of
  America}\ }\textbf {\bibinfo {volume} {97}},\ \bibinfo {pages} {2075}
  (\bibinfo {year} {2000})}\BibitemShut {NoStop}%
\bibitem [{\citenamefont {Ngoc}(2013)}]{Ngoc2013}%
  \BibitemOpen
  \bibfield  {author} {\bibinfo {author} {\bibfnamefont {P.~H.~A.}\
  \bibnamefont {Ngoc}},\ }\href {\doibase 10.1109/TAC.2012.2203031} {\bibfield
  {journal} {\bibinfo  {journal} {IEEE Transactions on Automatic Control}\
  }\textbf {\bibinfo {volume} {58}},\ \bibinfo {pages} {203} (\bibinfo {year}
  {2013})}\BibitemShut {NoStop}%
\bibitem [{\citenamefont {Chu}\ and\ \citenamefont {Barnes}(2016)}]{Chu2016}%
  \BibitemOpen
  \bibfield  {author} {\bibinfo {author} {\bibfnamefont {D.}~\bibnamefont
  {Chu}}\ and\ \bibinfo {author} {\bibfnamefont {D.~J.}\ \bibnamefont
  {Barnes}},\ }\href {\doibase 10.1038/srep25191} {\bibfield  {journal}
  {\bibinfo  {journal} {Scientific Reports}\ }\textbf {\bibinfo {volume} {6}},\
  \bibinfo {pages} {25191} (\bibinfo {year} {2016})}\BibitemShut {NoStop}%
\bibitem [{\citenamefont {Asmussen}\ \emph {et~al.}(1996)\citenamefont
  {Asmussen}, \citenamefont {Nerman},\ and\ \citenamefont
  {Olsson}}]{Asmussen1996}%
  \BibitemOpen
  \bibfield  {author} {\bibinfo {author} {\bibfnamefont {S.}~\bibnamefont
  {Asmussen}}, \bibinfo {author} {\bibfnamefont {O.}~\bibnamefont {Nerman}}, \
  and\ \bibinfo {author} {\bibfnamefont {M.}~\bibnamefont {Olsson}},\
  }\href@noop {} {\bibfield  {journal} {\bibinfo  {journal} {Scandinavian
  Journal of Statistics}\ }\textbf {\bibinfo {volume} {23}},\ \bibinfo {pages}
  {419} (\bibinfo {year} {1996})}\BibitemShut {NoStop}%
\bibitem [{\citenamefont {Cox}(1955)}]{Cox1955}%
  \BibitemOpen
  \bibfield  {author} {\bibinfo {author} {\bibfnamefont {D.~R.}\ \bibnamefont
  {Cox}},\ }\href {\doibase 10.1017/S0305004100030231} {\bibfield  {journal}
  {\bibinfo  {journal} {Mathematical Proceedings of the Cambridge Philosophical
  Society}\ }\textbf {\bibinfo {volume} {51}},\ \bibinfo {pages} {313}
  (\bibinfo {year} {1955})}\BibitemShut {NoStop}%
\bibitem [{\citenamefont {Ogura}\ and\ \citenamefont
  {Martin}(2014)}]{Ogura2013f}%
  \BibitemOpen
  \bibfield  {author} {\bibinfo {author} {\bibfnamefont {M.}~\bibnamefont
  {Ogura}}\ and\ \bibinfo {author} {\bibfnamefont {C.~F.}\ \bibnamefont
  {Martin}},\ }\href {\doibase 10.1137/130925177} {\bibfield  {journal}
  {\bibinfo  {journal} {SIAM Journal on Control and Optimization}\ }\textbf
  {\bibinfo {volume} {52}},\ \bibinfo {pages} {1809} (\bibinfo {year}
  {2014})}\BibitemShut {NoStop}%
\bibitem [{\citenamefont {Chichigina}\ \emph {et~al.}(2011)\citenamefont
  {Chichigina}, \citenamefont {Dubkov}, \citenamefont {Valenti},\ and\
  \citenamefont {Spagnolo}}]{Chichigina2011}%
  \BibitemOpen
  \bibfield  {author} {\bibinfo {author} {\bibfnamefont {O.~A.}\ \bibnamefont
  {Chichigina}}, \bibinfo {author} {\bibfnamefont {A.~A.}\ \bibnamefont
  {Dubkov}}, \bibinfo {author} {\bibfnamefont {D.}~\bibnamefont {Valenti}}, \
  and\ \bibinfo {author} {\bibfnamefont {B.}~\bibnamefont {Spagnolo}},\ }\href
  {\doibase 10.1103/PhysRevE.84.021134} {\bibfield  {journal} {\bibinfo
  {journal} {Physical Review E}\ }\textbf {\bibinfo {volume} {84}},\ \bibinfo
  {pages} {021134} (\bibinfo {year} {2011})}\BibitemShut {NoStop}%
\bibitem [{\citenamefont {Kargovsky}\ \emph {et~al.}(2015)\citenamefont
  {Kargovsky}, \citenamefont {Chichigina}, \citenamefont {Anashkina},
  \citenamefont {Valenti},\ and\ \citenamefont {Spagnolo}}]{Kargovsky2015}%
  \BibitemOpen
  \bibfield  {author} {\bibinfo {author} {\bibfnamefont {A.~V.}\ \bibnamefont
  {Kargovsky}}, \bibinfo {author} {\bibfnamefont {O.~A.}\ \bibnamefont
  {Chichigina}}, \bibinfo {author} {\bibfnamefont {E.~I.}\ \bibnamefont
  {Anashkina}}, \bibinfo {author} {\bibfnamefont {D.}~\bibnamefont {Valenti}},
  \ and\ \bibinfo {author} {\bibfnamefont {B.}~\bibnamefont {Spagnolo}},\
  }\href {\doibase 10.1103/PhysRevE.92.042140} {\bibfield  {journal} {\bibinfo
  {journal} {Physical Review E}\ }\textbf {\bibinfo {volume} {92}},\ \bibinfo
  {pages} {042140} (\bibinfo {year} {2015})}\BibitemShut {NoStop}%
\bibitem [{\citenamefont {Borkar}(1995)}]{Borkar1995}%
  \BibitemOpen
  \bibfield  {author} {\bibinfo {author} {\bibfnamefont {V.~S.}\ \bibnamefont
  {Borkar}},\ }\href@noop {} {\emph {\bibinfo {title} {{Probability Theory}}}}\
  (\bibinfo  {publisher} {Springer-Verlag New York},\ \bibinfo {year}
  {1995})\BibitemShut {NoStop}%
\end{thebibliography}
\end{document}